\documentclass[12pt]{article}
\pdfoutput=1
\usepackage{putex}
\usepackage{feyn}
\usepackage[vcentermath]{youngtab}

\usepackage{graphicx}
\usepackage{epstopdf}
\usepackage{enumerate}
\usepackage{cite}
\usepackage{tensor}
\usepackage{slashed}
\usepackage{feynmf}

\usepackage{hyperref}

\numberwithin{equation}{section}

\newcommand{\HH}{\mathbb{H}}
\newcommand{\OO}{\mathbb{O}}
\newcommand{\EE}{\mathbb{E}}

\newcommand{\abs}[1]{\left\lvert #1 \right\rvert}

\newcommand {\be} {\begin {equation}}
\newcommand {\ee} {\end {equation}}

\newcommand {\bes} {\begin {equation*}}
\newcommand {\ees} {\end {equation*}}

\newcommand{\es}[2] {\begin{equation} \label{#1} \begin{split} #2 \end{split} \end{equation}}

\newcommand{\Z}{\mathbb{Z}}

\newcommand{\R}{\mathbb{R}}

\newcommand{\beq}{\begin{equation}}
\newcommand{\eeq}{\end{equation}}

\def\be{ \begin{equation} }
\def\ee{ \end{equation} }

\def\res{\mathop{\text{Res}}}

\begin{document}

\preprint{MIT-CTP-4471}

\institution{PU}{Department of Physics, Princeton University, Princeton, NJ 08544}
\institution{PCTS}{Princeton Center for Theoretical Science, Princeton University, Princeton, NJ 08544}
\institution{MIT}{Center for Theoretical Physics, Massachusetts Institute of Technology, Cambridge, MA 02139}

\title{
AdS Description of Induced Higher-Spin Gauge Theory
}

\authors{Simone Giombi,\worksat{\PU} Igor R.~Klebanov,\worksat{\PU,\PCTS} Silviu S.~Pufu,\worksat{\MIT} Benjamin R.~Safdi,\worksat{\PU}\\[10pt] and
Grigory Tarnopolsky\worksat{\PU}
}

\abstract{
We study deformations of three-dimensional large $N$ CFTs by double-trace operators constructed from spin $s$ single-trace operators of dimension $\Delta$. These theories possess UV fixed points, and we calculate the change of the 3-sphere free energy $\delta F= F_{\rm UV}- F_{\rm IR}$.  To describe the UV fixed point using the dual AdS$_4$ space we modify the boundary conditions on the spin $s$ field in the bulk; this approach produces $\delta F$ in agreement with the field theory calculations.  If the spin $s$ operator is a conserved current, then the fixed point is described by an induced parity invariant conformal spin $s$ gauge theory. The low spin examples are QED$_3$ ($s=1$) and the 3-d induced conformal gravity ($s=2$). When the original CFT is that of $N$ conformal complex scalar or fermion fields, the $U(N)$ singlet sector of the induced 3-d gauge theory is dual to Vasiliev's theory in AdS$_4$ with alternate boundary conditions on the spin $s$ massless gauge field. We test this correspondence by calculating the leading term in $\delta F$ for large $N$. We show that the coefficient of $\frac 1 2 \log N$ in $\delta F$ is equal to the number of spin $s-1$ gauge parameters that act trivially on the spin $s$ gauge field.  We discuss generalizations of these results to 3-d gauge theories including Chern-Simons terms and to theories where $s$ is half-integer.  We also argue that the Weyl anomaly $a$-coefficients of conformal spin $s$ theories in even dimensions $d$, such as that of the Weyl-squared gravity in $d=4$, can be efficiently calculated using massless spin $s$ fields in AdS$_{d+1}$ with alternate boundary conditions. Using this method we derive a simple formula for the Weyl anomaly $a$-coefficients of the $d=4$ Fradkin-Tseytlin conformal higher-spin gauge fields. Similarly, using alternate boundary conditions in AdS$_3$ we reproduce the well-known central charge $c=-26$ of the $bc$ ghosts in 2-d gravity, as well as its higher-spin generalizations.}

\date{}
\maketitle

\tableofcontents

\section{Introduction and summary}

A Conformal Field Theory (CFT) in $d$ dimensions is dual to a gravitational theory in AdS$_{d+1}$ endowed with a particular choice of boundary conditions
\cite{Maldacena:1997re,Gubser:1998bc,Witten:1998qj}. For example, a local scalar operator ${\cal O}(x^\mu)$ with dimension $\Delta$ is dual to a scalar field $\Phi (z, x^\mu)$
that behaves as $z^\Delta$ near the AdS boundary. The possible values of $\Delta$ are determined by the mass of the scalar field in the bulk:
\es{opdim}{
\Delta_\pm = \frac d 2 \pm \sqrt { \left ({\frac d 2}\right )^2 + m^2 }
\ ,}
where the AdS radius has been set to $1$. The dimension $\Delta_-$ is allowed only in the range $-(d/2)^2 < m^2 < -(d/2)^2+1$ \cite{Breitenlohner:1982jf,Klebanov:1999tb}; using it for greater
values of $m^2$ results in an operator dimension that violates the unitarity bound. An RG flow from a large $N$ CFT where the operator ${\cal O}$ has dimension
$\Delta_-$ to another CFT where it has dimension $\Delta_+$ takes place when the double-trace operator ${\cal O}^2$ is added to the action
\cite{Witten:2001ua,Gubser:2002vv}. The effect of this flow on the partition function of the Euclidean CFT on the $d$-dimensional sphere has been studied in
a number of papers \cite{Gubser:2002zh,Gubser:2002vv,Hartman:2006dy,Diaz:2007an,Klebanov:2011gs}.

These results have interesting applications to
AdS$_4$/CFT$_3$ dualities involving Vasiliev's interacting higher-spin gauge theories in AdS$_4$ \cite{Vasiliev:1990en,Vasiliev:1992av,Vasiliev:1995dn,Vasiliev:1999ba}. These theories have been conjectured to be dual to 3-d CFTs such as the critical $O(N)$ model \cite{Klebanov:2002ja},
or the Gross-Neveu model \cite{Sezgin:2003pt,Leigh:2003gk}, or various large $N$ Chern-Simons theories coupled to conformal matter in the fundamental representation of the gauge group \cite{Aharony:2011jz,Giombi:2011kc}. Such AdS/CFT dualities are often called ``vectorial'' because the dynamical fields in the CFT are $N$-vectors rather than $N\times N$ matrices.
In particular, the scalar $O(N)$ model has been conjectured \cite{Klebanov:2002ja} to be dual to the
minimal type-A Vasiliev theory containing gauge fields of all even spin in AdS$_4$, while the Gross-Neveu model has been conjectured \cite{Sezgin:2003pt,Leigh:2003gk} to be dual to the minimal type-B Vasiliev theory.\footnote{An important distinction between the type A and B parity invariant Vasiliev theories is that in the former the scalar field has positive
 parity, while in the latter it has negative parity \cite{Vasiliev:1992av,Sezgin:2003pt}.} Considerable evidence has been accumulated in favor of the vectorial AdS$_4$/CFT$_3$ dualities \cite{Giombi:2009wh,Giombi:2010vg,Giombi:2011ya,Maldacena:2011jn,Maldacena:2012sf,Colombo:2012jx,Didenko:2012tv,Gelfond:2013xt, Didenko:2013bj}, and we will make further use of them in this paper.

The possibility of two different conformally invariant AdS boundary conditions extends in an interesting way to fields of spin $s>0$.
For example, to a spin $1$ conserved $U(1)$ current $J_\mu$ in a 3-dimensional CFT there corresponds a massless gauge field $A_\mu$ in AdS$_4$ with the boundary condition
that the magnetic field $F_{ij}$ vanishes at the AdS boundary $z=0$. If instead the electric field $F_{iz}$ is required to vanish at the boundary, then
the $U(1)$ symmetry of the CFT becomes gauged \cite{Witten:2003ya}. These facts have applications to the versions of Vasiliev theory that contain gauge fields of all
integer spin in AdS$_4$. The type A such model is dual to the $U(N)$ symmetric 3-d CFT of $N$ complex scalar fields \cite{Klebanov:2002ja}, while the type B model is dual to the theory of $N$ Dirac fermions \cite{Sezgin:2003pt,Leigh:2003gk}. The ability to change the boundary conditions for the spin $1$ field
makes it plausible \cite{Giombi:2012ms} that the type A or B
Vasiliev theory in AdS$_4$ with the electric boundary condition on the spin $1$ field is dual to 3-dimensional CFTs
where the $U(1)$ gauge field is coupled to a large number $N$ of conformally invariant complex scalar or fermion fields, i.e.~the
3-dimensional ``induced'' QED \cite{Appelquist:1981vg} restricted to the $SU(N)$ singlet sector. A more general, mixed boundary condition on the $U(1)$ gauge field in AdS$_4$ results in addition
of the Chern-Simons term for the dynamical $U(1)$ gauge field in QED$_3$ \cite{Witten:2003ya}. There is an $SL(2,\Z)$ action on the resulting set of
3-d CFTs \cite{Witten:2003ya}. The possibility of imposing modified boundary conditions on spins $s\le 1$ in Vasiliev's theory was also used in \cite{Chang:2012kt} in constructing higher-spin duals of various supersymmetric Chern-Simons matter theories. Besides considering the $U(1)$ symmetries Ref.~\cite{Chang:2012kt} also considered gauging
non-abelian symmetries. Non-abelian gauge fields can appear in supergravity as well as in Vasiliev theory; with standard boundary conditions they correspond to non-abelian global symmetries in the dual field theory. Changing the boundary conditions in AdS$_{d+1}$ is expected to lead to a non-abelian induced gauge theory
in $d$ dimensions.

Another very interesting special case is $s=2$. Modifying the boundary condition for the graviton in AdS$_4$ makes the metric fluctuating also in the
dual boundary theory \cite{Leigh:2003ez,Compere:2008us}. The resulting 3-d theory then describes a Weyl invariant gravity induced by coupling to conformal matter.
The effective action for this theory was explored at the quadratic order for gravitons in \cite{Leigh:2003ez}. A further study of the modified boundary
conditions in AdS$_4$ indicated that the correspondence with 3-d induced gravity works at the full non-linear level \cite{Compere:2008us}. Furthermore, the conformal graviton spectrum around flat space was found in \cite{Compere:2008us} to be free of ghost-like modes for all odd $d$, suggesting that these induced theories are unitary at least in perturbation theory (on the other hand, in even $d$ there are ghosts, as familiar in the case of $d=4$ Weyl gravity \cite{Fradkin:1985am}).
Using these ideas, we will conjecture, for example, that modifying the graviton boundary conditions in Vasiliev's minimal type A theory makes it dual
to the $O(N)$ singlet sector of the Weyl invariant 3-d gravity coupled to $N$ conformal scalar fields $\phi^i$, $i=1, \ldots, N$. The path integral for this theory is
\begin{eqnarray}
\label{Weylgrav}
Z_{\text{3-d gravity}} & = &\int {[D g_{\mu\nu}] [D\phi^i]\over {\rm Vol}(\text{Diff}) {\rm Vol} (\text{Weyl})} e^{-S}\ , \\
S &=& \int d^3 x \sqrt{g} \left ( g^{\mu \nu} \partial_\mu \phi^i \partial_\nu \phi^i + \frac 1 8 R (\phi^i)^2 \right )
\ .
\end{eqnarray}
Similarly, it is plausible that the minimal type B Vasiliev theory with modified graviton boundary conditions is dual to the $O(N)$ singlet sector of the Weyl invariant 3-d gravity coupled to $N$
massless fermions. As for the $s=1$ case, for $s=2$ there is a possibility of mixed parity-violating boundary conditions in AdS$_4$ \cite{Leigh:2003ez, Compere:2008us,deHaro:2007fg,deHaro:2008gp},
which correspond to adding
to the 3-d action the gravitational Chern-Simons term $i \kappa \int \tr (\omega \wedge d\omega + \frac 2 3 \omega^3)$ \cite{Deser:1982vy,Deser:1981wh}. Similarly,
the ${\cal N}=8$ superconformal gravity coupled to the BLG/ABJM theory was studied in \cite{Gran:2008qx, Nilsson:2012ky, Nilsson:2013fya}. The crucial role of alternate boundary conditions in AdS$_4$ was noted there as well. 

In analogy with the above discussions, it is possible to modify the AdS$_4$ boundary conditions for higher-spin fields with $s>2$.
  This modification results in gauging the corresponding
higher-spin symmetries in the 3-d boundary theory,\footnote{
  One motivation for studying the theories where some of the currents are gauged, which was stressed in \cite{Vasiliev:2012vf}, is that they do not obey the
theorem of \cite{Maldacena:2011jn}. This theorem requires theories with exactly conserved higher spin currents to be free. However, when some of the
 currents are gauged the remaining ones are not conserved; therefore, the
theorem of \cite{Maldacena:2011jn} does not apply. For example, the 3-d QED coupled to $N$ flavors is obviously not a free theory, even when $N$ is large. The theory obtained by gauging the whole set of HS currents also does not obey the theorem of \cite{Maldacena:2011jn}, being a higher spin gauge theory (in particular including gravity), while \cite{Maldacena:2011jn} assumes a CFT with global HS symmetries and corresponding exactly conserved currents.}
as was proposed some time ago at the level of the linearized approximation \cite{Leigh:2003ez} (see also \cite{Metsaev:2009ym})
and studied more recently in the context of the fully non-linear Vasiliev higher-spin theory \cite{Vasiliev:2012vf}.
The non-linearities have the important effect that, when an $s>2$ current is gauged, one may need to gauge all remaining currents too.\footnote{
We thank M. Vasiliev for stressing this to us.} In that case, the 3-d dual of a minimal Vasiliev theory in AdS$_4$ is expected to be a Weyl invariant theory of gauge fields of all even spins induced by the
coupling to $N$ conformal scalar or fermion fields. On the other hand,
the gauged $s=1$ and $s=2$ examples discussed above
do not require gauging higher-spin symmetries, because the non-linear gauge transformations for spin $s\leq 2$ form a closed subalgebra of the higher-spin algebra.  The 3-d theory where currents of all spin are gauged is clearly more complicated than either $3$-dimensional QED or the induced gravity theory in \eqref{Weylgrav}. Such an induced higher-spin gauge theory was studied in \cite{Bekaert:2010ky}, and some progress has been recently made using twistor space techniques in the unfolded formulation \cite{Vasiliev:2012vf}. It is also interesting to ask if a truncation of this 3-d theory to a finite number of higher-spins is possible.

In this paper we will subject these Anti-de Sitter/Induced Gauge Theory (AdS/IGT) correspondences to some new tests in the regime where $N$ is very large;  in this limit the Vasiliev theories in AdS$_4$ become weakly coupled while the path integrals in the 3-d theory can be studied semi-classically. We will calculate the change in the 3-sphere free energy $F=- \log \abs{Z_{S^3}}$ produced by the gauging of a symmetry with $s\geq 1$. We will then show that this change agrees with the corresponding calculation in Euclidean AdS$_4$, which uses modified boundary conditions
for a spin $s$ gauge field. In fact, in QED$_3$ coupled to $N$ conformal scalar or fermion fields the 3-sphere free energy was
studied in \cite{Klebanov:2011td} with the result
$F_{\rm QED}- F_{\rm free}  =
\frac{1}{2} \log N + O(N^0)$. We will show that for the gauging of spin $s$ current this expression generalizes to
\es{lNdivergence}{
F^{(s)}_{\rm gauged}- F^{(s)}_{\rm free} = {(4 s^2-1) s\over 6} \log N + O(N^0)\ .
}
As we will discuss in section \ref{conscurr},
the coefficient of $\frac 1 2 \log N$ is the number of spin $s-1$ conformal Killing tensors (equivalently, these are the conformal higher-spin currents which were found in \cite{Konstein:2000bi} following \cite{Lopatin:1987hz}).  Each such tensor corresponds to a
missing gauge invariance (a zero mode of the operator ${\cal O}_g$ defined in~\eqref{Og} that takes a rank $s-1$ traceless symmetric tensor to a pure gauge mode of a spin $s$ gauge field) in the $3$-dimensional theory of the spin $s$ gauge field. These tensors transform in the $[s-1, s-1]$ irreducible representation
of the conformal group $SO(4,1)$ (its Young tableaux has two rows of length $s-1$) \cite{Vasiliev:2001wa,1091.53020}. The AdS/CFT correspondence relates a conformal Killing tensor in $d$
dimensions to a traceless Killing tensor in AdS$_{d+1}$ \cite{Mikhailov:2002bp}. In section \ref{masslessAdS} we will study this relation in detail with special emphasis on the AdS boundary behavior of the Killing tensors.

In addition to studying the gauging of conserved higher-spin currents, we will study the closely related problem of deforming a 3-d CFT by
a double-trace operator $J_{\mu_1 \mu_2 \dots \mu_s} J^{\mu_1 \mu_2 \dots \mu_s}$, where
the spin $s$ single-trace operator $J_{\mu_1 \mu_2 \dots \mu_s}$ has dimension $\Delta$. If
$\Delta > 3/2$,
then the double-trace operator is irrelevant; such irrelevant deformations were discussed for $s\geq 1$ in \cite{Leigh:2003ez}. For large $N$ it is possible to show that the deformed theory possesses
a UV fixed point where the spin $s$ operator has dimension $\Delta_-= 3 - \Delta + O(1/N)$.
In this case, we will find using both the 3-d field theory and AdS$_4$ calculations that
\es{GenFormula}{
\delta F^{(s)}_\Delta \equiv F^{(s)}_{\rm UV}- F^{(s)}_{\rm IR} = {(2 \, s + 1) \pi \over 6} \int_{3/2}^\Delta \big( x - {3 \over 2} \big) (x + s - 1)(x - s - 2) \cot (\pi x) \,.
}
For spin $s\geq 1$, $\Delta_-$ cannot satisfy the unitarity bound $\Delta^{(s)}\geq s+1$. The only cases where unitarity appears to be restored is when the spin $s$ current is conserved
and has $\Delta=s+1$; then $\Delta_-$ is the dimension of the dual spin $s$ gauge field, which is not a gauge invariant operator, so there is no obvious issue with unitarity.

While in odd dimensions $d$ the parity invariant conformal higher-spin gauge theories have induced non-local actions, in even $d$ there are theories that are {\it local} and Weyl invariant for any spin $s$ (these local actions are the coefficients of the induced logarithmically divergent terms \cite{Liu:1998bu,Tseytlin:2002gz, Segal:2002gd, Compere:2008us}). For example, in $d=4$ they are  the free Maxwell theory ($s=1$), the conformal gravity ($s=2$) \cite{Fradkin:1983tg}, and their Fradkin-Tseytlin higher-spin generalizations \cite{Fradkin:1985am}.
These conformal higher-spin theories have actions involving more than two derivatives in contrast with the two-derivative quadratic
Fronsdal actions \cite{Fronsdal:1978rb}. This is evident already for the $s=2$ conformal theory whose action is the square of the Weyl tensor.
The role of the Weyl-squared gravity in the AdS/CFT correspondence has been explored for some time \cite{Liu:1998bu,Compere:2008us}.  A relation between conformal $d=4$ higher-spin theories and massless higher-spin theories in AdS$_5$ was proposed in \cite{Tseytlin:2002gz, Segal:2002gd}.
Our approach of using alternate boundary conditions for massless spin $s$ gauge fields in Euclidean AdS$_{d+1}$ indeed relates them to conformal spin $s$
gauge fields on $S^d$. As an application of these ideas,
in section \ref{Weylsec} we will demonstrate that the massless spin $s$ fields in AdS$_{d+1}$ endowed with alternate boundary conditions provide an efficient way for
calculating the Weyl anomaly $a$-coefficients of conformal spin $s$ theories in even $d$. In particular, we will reproduce the Weyl anomaly $a$-coefficient of the $d=4$
conformal gravity \cite{Fradkin:1983tg,Fradkin:1985am} and conjecture a formula generalizing it to all conformal 4-d gauge theories of integer spin $s>0$:
\es{higheranom}{
\qquad a_s = {s^2 \over 180} (1+s)^2 [ 3 + 14 s (1 + s) ] \, .
}

Similarly, we may consider higher-spin theories in AdS$_3$ \cite{Blencowe:1988gj,Bergshoeff:1989ns,Prokushkin:1998bq} whose dual $d=2$ CFTs have $W$ symmetries
\cite{Henneaux:2010xg,Campoleoni:2010zq,Gaberdiel:2010ar,Gaberdiel:2010pz,Gaberdiel:2011wb,Gaberdiel:2012uj}. Changing the boundary conditions in the bulk corresponds to gauging these symmetries. From the one-loop determinants of graviton and higher-spin gauge fields with alternate boundary conditions in AdS$_3$, we reproduce the well-known central charge $c=-26$ of the $bc$ ghosts in 2-d gravity \cite{Polyakov:1981rd}, as well as its higher-spin generalization
\cite{Pope:1991uz}: $c_s=-2(1+6s(s-1))$ .

\section{Double-trace deformations with higher-spin operators}

We start by analyzing the double-trace deformations with $s \geq 1$ in the case where the single-trace spin $s$ operator has dimension $\Delta \neq s + 1$.  As remarked in the introduction, these deformations are somewhat less desirable than those with $\Delta = s+1$ due to the appearance of operators that violate the unitarity bound.  Nevertheless, the theories with $\Delta \neq s+1$ are still interesting conceptually, and they are somewhat simpler computationally because we do not have to worry about gauge invariance.  As a consequence of this fact---and we will show this in detail in the following sections---the difference in free energies $\delta F_\Delta^{(s)}$ is order $N^0$ when $\Delta \neq s + 1$, while it is order $\log N$ when $\Delta = s + 1$, as advertised in~\eqref{GenFormula} and~\eqref{lNdivergence}.  In this section we begin with the cases $\Delta \neq s + 1$ and use field theoretic arguments to demonstrate~\eqref{GenFormula} for small values of $s$.  In section \ref{conscurr} we then discuss the implications of gauge invariance when $\Delta = s+1$.

\begin{figure}[htb]
  \leavevmode
\begin{center}$
\begin{array}{cc}
\scalebox{.55}{\includegraphics{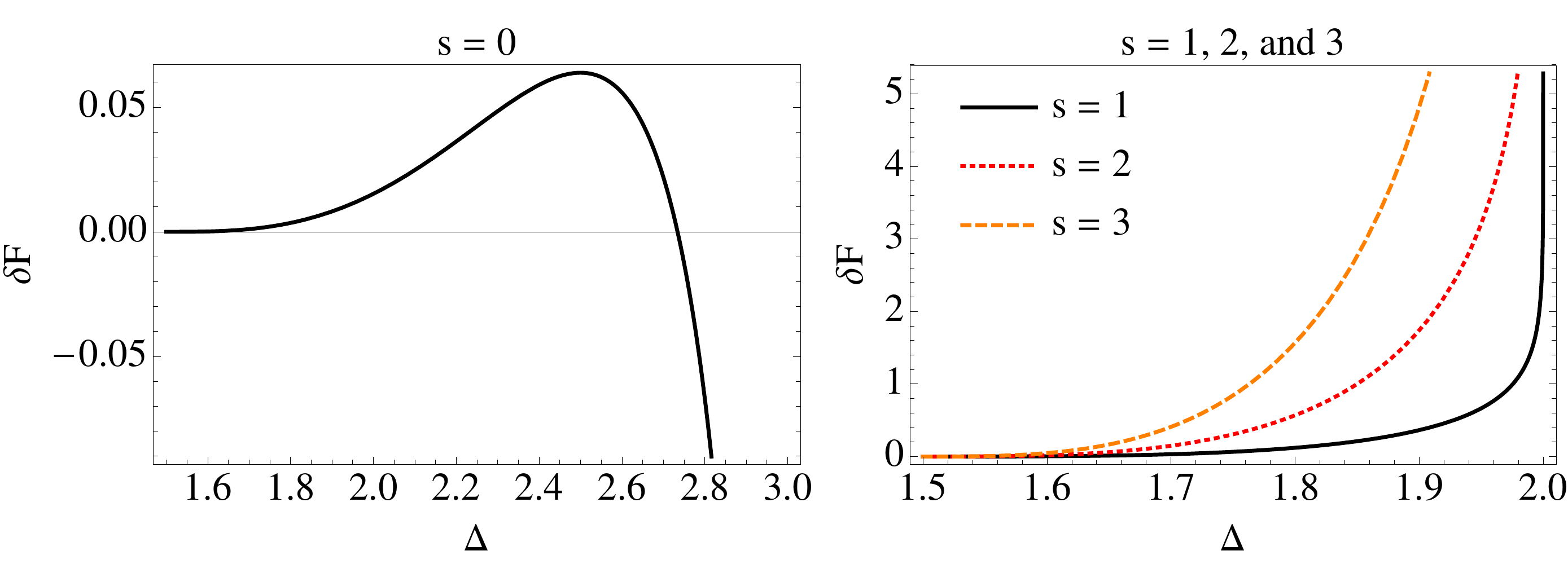}} \\
 \end{array}$
\end{center}
\caption{$\delta F_\Delta^{(s)}$ plotted as a function of $\Delta$ for $s =0$, 1, 2, and 3.  When $s = 0$ this quantity is required by the $F$-theorem to be positive for $3/2 < \Delta < 5/2$, but outside of this range and also for higher-spin, the $F$-theorem does not apply since one or both of the fixed points is non-unitary.  The exception is when $\Delta = s+1$, since in this case the naive unitarity arguments are not valid.   }
\label{images}
\end{figure}

 Before turning to the calculation, however, we mention two interesting features of the result in~\eqref{GenFormula}.  The first observation is that $\delta F_\Delta^{(s)}$ is positive for $3/2 < \Delta <2$ for all $s$.  When $s = 0$ this is required by the $F$-theorem
 \cite{Jafferis:2011zi,Klebanov:2011gs,Myers:2010tj,Casini:2011kv,Casini:2012ei}---in fact, in that case $\delta F$ must be positive when $3/2 < \Delta < 5/2$.  For $\Delta>5/2$ the UV fixed point is non-unitary because $\Delta_- < 1/2$, and the $F$-theorem is not required to hold. Indeed, for $\Delta$ greater than $\approx 2.73423$ there is a region where $F_{\rm UV}- F_{\rm IR}$ is negative, as illustrated in figure~\ref{images}.\footnote{Similarly, the Zamolodchikov $c$-theorem
 \cite{Zamolodchikov:1986gt} is not applicable to non-unitarity theories. For explicit violations of the $c$-theorem
 in non-unitary theories see, for example, \cite{Freedman:1990tg,Mende:1989md}.}
Similarly, when $s \geq 1$ one of the fixed points is always non-unitary, and so the $F$-theorem does not require $\delta F$ to be positive.  It is therefore interesting
 that $\delta F_\Delta^{(s)}$ is always positive for $3/2 < \Delta <2$, but the significance of this observation is unclear.

 The second observation, which is also illustrated in figure~\ref{images}, is that $\delta F_\Delta^{(s)}$ diverges logarithmically as $\Delta \to 2$ when $s \geq 1$.  Furthermore, if we take $\Delta = s + 1 - \epsilon$,  where $\epsilon \ll 1$, and concentrate on the contribution of the upper integration limit in~\eqref{GenFormula}, then we find
\es{epsdivergence}{
\delta F_\Delta^{(s)} = -{(4 s^2-1) s\over 6} \log \epsilon + O(\epsilon^0)\ .
}
This result shows, in some sense, how the result in~\eqref{lNdivergence}, which is valid strictly when $\Delta = 1 +s$, emerges from the case of more general double-trace
deformation. The conclusion is that gauging a symmetry in a large $N$ CFT makes $\delta F$ logarithmically large.

\subsection{General strategy}

The RG flow we are considering may be constructed explicitly as follows. Let $S_0$ be the action of a large $N$ CFT defined on a conformally flat background with metric $g_{\mu\nu}$. We perturb $S_0$ by the irrelevant deformation proportional to the double-trace operator $J^2$ to obtain the action
\es{actionMod}{
S = S_0 + {\lambda_0 \over 2} \int d^3x \, \sqrt g J_{\mu_1 \mu_2 \dots \mu_s}(x) J^{\mu_1 \mu_2 \dots \mu_s}(x) \, ,
}
where $J_{\mu_1 \mu_2 \dots \mu_s}$ is a symmetric traceless tensor.
This theory has a UV fixed point where $J_{\mu_1 \mu_2 \dots \mu_s}$ has dimension $\Delta_-  = 3 - \Delta+ O(1/N)$. To demonstrate this, we use the
Hubbard-Stratonovich transformation to write the action with the help of a spin $s$ auxiliary field $h_{\mu_1 \mu_2 \dots \mu_s}$:
 \es{SHubbard}{
  S = S_0 - \int d^3 x\, \sqrt{g(x)} \left[h_{\mu_1 \ldots \mu_s}(x) J^{\mu_1 \ldots \mu_s}(x)
  +\frac{1}{2 \lambda_0} h_{\mu_1 \ldots \mu_s} h^{\mu_1 \ldots \mu_s}
  \right] \,.
 }
A study of the induced action for $h_{\mu_1 \mu_2 \dots \mu_s}$ shows that the last term is negligible at the UV fixed point \cite{Leigh:2003ez}. When the current $J_{\mu_1 \mu_2 \dots \mu_s}$ is conserved, the auxiliary field $h_{\mu_1 \mu_2 \dots \mu_s}$ assumes the role of a spin $s$ gauge field.

One can evaluate the ratio $Z/Z_0$ of the partition functions corresponding to $S$ and $S_0$ perturbatively in $1/N$ as follows.  Integrating out the fields that appear in the undeformed action $S_0$, one can write the partition function of the deformed theory \eqref{SHubbard} as
 \es{PartInteracting}{
  Z/Z_0 = \int D h_{\mu_1 \ldots \mu_s} \left\langle \exp \left(\int d^3x \, \sqrt{g(x)} h_{\mu_1 \ldots \mu_s}(x) J^{\mu_1 \ldots \mu_s}(x)\right)\right\rangle_0 \,,
 }
where on the right-hand side the expectation value is computed with the measure $\exp[-S_0]$.
Expanding the exponential and using the fact that $\langle  J_{\mu_1 \ldots \mu_s}(x) \rangle_0 = 0$, as appropriate for a CFT on a conformally flat space, one obtains
 \es{PartEffAction}{
  Z = Z_0 \int D h_{\mu_1 \ldots \mu_s} e^{-S_\text{eff}[h_{\mu_1 \ldots \mu_s}]} \,,
 }
where the effective action for the auxiliary field is to quadratic order given by
 \es{Seff}{
  S_\text{eff}= -\frac 12 \int d^3x \, d^3y\, \sqrt{g(x)} \sqrt{g(y)} \, h_{\mu_1 \ldots \mu_s}(x)  h_{\nu_1 \ldots \nu_s}(y)
   \langle J^{\mu_1 \ldots \mu_s}(x) J^{\nu_1 \ldots \nu_s}(y) \rangle^\text{conn}_0 + \ldots \,.
 }
The expansion in \eqref{Seff} is given in terms of connected correlators of the spin $s$ operator, which are all assumed to be $O(N)$.  At large $N$ the typical fluctuations of $h_{\mu_1 \ldots \mu_s}$ are $O(N^{-1/2})$, and therefore the contributions to the partition function of the higher order terms in $h_{\mu_1 \ldots \mu_s}$, that were not exhibited in \eqref{Seff}, become negligible.  The functional integral \eqref{PartEffAction} can then be evaluated in the saddle-point approximation:
 \es{ZSaddle}{
  Z \approx Z_0 (\det K)^{-1/2} \,,
 }
where the operator $K$ given as an integration kernel can be expressed as
 \es{Kernel}{
  K_{\mu_1\ldots \mu_s; \nu_1 \ldots \nu_s}(x, y) = -\langle J_{\mu_1 \ldots \mu_s}(x) J_{\nu_1 \ldots \nu_s}(y) \rangle^\text{conn}_0 \,.
 }
The expression \eqref{ZSaddle} is valid on any conformally flat space.

Specializing to the case where the background metric is that of the unit $S^3$, \eqref{ZSaddle} implies
 \es{deltaF}{
  \delta F_\Delta^{(s)} = -\log \abs{\frac{Z}{Z_0}} = \frac 12 \tr \log K + O(1/N) \,.
 }
To calculate $\delta F_\Delta^{(s)}$ one would therefore need to sum the logarithms of the eigenvalues of the kernel $K$ on $S^3$ weighted by their multiplicities.

An explicit formula for $K$ can be written down most easily if we parameterize $S^3$ through the stereographic projection from $\R^3$.  In other words, let us introduce the metric
 \es{S3Metric}{
  ds^2_{S^3} = \frac{4}{\left(1 + \abs{x}^2\right)^2} \left[ (dx^1)^2 + (dx^2)^2 + (dx^3)^2 \right] \,,
 }
as well as the frame
 \es{S3Frame}{
  e^i = \frac{2}{1 + \abs{x}^2}  dx^i \,.
 }
In this frame, the kernel \eqref{Kernel} is constrained by conformal invariance to be\footnote{Frame indices are raised and lowered with the flat metric.}
 \es{KS3}{
  K_{i_1\ldots i_s}{}^{j_1 \ldots j_s}(x, y)
   = N \,C \left( \frac{\left( 1 + \abs{x}^2 \right) \left( 1 + \abs{y}^2 \right)}{4 \abs{x-y}^2} \right)^\Delta I_{(i_1}{}^{(j_1} I_{i_2}{}^{j_2} \cdots I_{i_s)}{}^{j_s)} \,,
 }
where $C$ is an $N$-independent normalization constant, and
 \es{IDef}{
  I^{ij} \equiv \delta^{ij} - 2 \frac{(x^i - y^i) (x^j-y^j)}{\abs{x-y}^2} \,.
 }
In \eqref{KS3}, the symmetrizations are performed with total weight one and include the removal of all the traces.  Importantly, the kernel $K$ is linear in $N$.

\section{Explicit field theory calculations}
\subsection{Symmetric traceless tensor harmonics on $S^3$}
\label{HARMONICS}

The eigenvalues of $K$ can be found with the help of rotational symmetry on $S^3$; the eigenfunctions of $K$ must be symmetric traceless tensor harmonics on $S^3$.  For spin $0$, these harmonics are the usual spherical harmonics on $S^3$ which transform as the $({\bf n}, {\bf n})$ irreps\footnote{We write the spin $j$ representation of $SU(2)$ as ${\bf 2 j + 1}$.} of the isometry group $SU(2)_L \times SU(2)_R$---they are just traceless symmetric polynomials in the standard embedding coordinates of $S^3$ into $\R^4$.   The space of normalizable functions on $S^3$ therefore decomposes under $SO(4)$ as
 \es{IrredScalar}{
  \bigoplus_{n=1}^\infty ({\bf n}, {\bf n} ) \,.
 }
For every positive integer $n$, there are $n^2$ scalar harmonics, which we denote by $Y_{n \ell m}(x)$, with $0 \leq \ell < n$ and $\abs{m} \leq \ell$.  Explicit expressions for these scalar harmonics are given in~\eqref{Ynlm}.

For spin $s$, the space of rank $s$ symmetric traceless tensors on $S^3$ decomposes under $SO(4)$ as
 \es{Irred}{
  \bigoplus_{n=s+1}^\infty \bigoplus_{s' = -s}^s ({\bf n+ s'}, {\bf n - s'} ) \,.
 }
In other words, there are $2s+1$ towers of modes indexed by $s'$, where there are $n^2 - s'^2$ modes in each tower, with $n > s$.  We denote these harmonics by $\HH_{\mu_1 \ldots \mu_s}^{s', n\ell m}(x)$, with $s' \leq \ell < n$ and $-\ell \leq m \leq \ell$.  Explicit expressions for $s \leq 3$ are given in Appendix~\ref{HARMONICSAPP}.

The reason for the decomposition \eqref{Irred} is easy to state.  Starting with the three $SU(2)_L$ Killing vectors (or the corresponding one-forms obtained by lowering indices with the metric), one can construct rank-$s$ traceless symmetric tensors by taking traceless symmetric tensor products of these Killing vectors.  Angular momentum addition guarantees that these tensors transform as $({\bf 2s+1}, {\bf 1})$ under $SU(2)_L \times SU(2)_R$.  The most general rank-$s$ traceless symmetric tensor on $S^3$ is a linear combination of these $({\bf 2s+1}, {\bf 1})$ tensors with coefficients that depend on position.  These coefficients are functions on $S^3$, so they can be expanded in the basis of scalar spherical harmonics, which as mentioned above transform as $({\bf n}, {\bf n})$ under $SO(4)$.  The traceless symmetric tensors therefore transform as the tensor sum of products $({\bf n}, {\bf n}) \otimes ({\bf 2s+1}, {\bf 1})$ over all $n \geq 1 $.  This description yields \eqref{Irred} after a shift in $n$.

All the harmonics in a given irreducible representation of $SO(4)$ are eigenfunctions of $K$ corresponding to the same eigenvalue.  Let $k_{n, s'}$ be the eigenvalue corresponding to each term in \eqref{Irred}:
 \es{Evalue}{
  \int d^3y\, \sqrt{g(y)} K_{\mu_1\ldots \mu_s}{}^{\nu_1 \ldots \nu_s}(x, y)\, \HH_{\nu_1 \ldots \nu_s}^{s',n\ell m}(y)
   = k_{n, s'} \HH_{\mu_1 \ldots \mu_s}^{s',n\ell m}(x) \,.
 }
Then
 \es{deltaFGen}{
  \delta F_\Delta^{(s)} = \frac 12 \sum_{n=s+1}^\infty  \sum_{s'=-s}^s (n^2 - s'^2) \log k_{n, s'} \,.
 }
Because the kernel \eqref{KS3} is invariant under the $\Z_2$ reflection symmetry that exchanges $SU(2)_L$ with $SU(2)_R$, we must have $k_{n, s'} = k_{n, -s'}$.  Since the eigenvalue $k_{n, s'}$ doesn't depend on the quantum numbers $\ell$ and $m$, we can write
 \es{knGeneral}{
  k_{n, s'} = \frac{1}{n^2 - s'^2} \sum_{\ell, m} \int d^3x \, d^3y\, \sqrt{g(x)} \sqrt{g(y)}\, \HH_{\mu_1 \ldots \mu_s}^{s',n\ell m}(x)^* \,
    K^{\mu_1\ldots \mu_s; \nu_1 \ldots \nu_s}(x, y)\, \HH_{\nu_1 \ldots \nu_s}^{s',n\ell m}(y) \,.
 }
The average over all the states in a given irreducible representation of $SO(4)$ makes the product $\HH(x)^* K(x, y) \HH(y)$ depend only on the relative angle between $x$ and $y$.  One can then perform five of the six integrals in \eqref{knGeneral}, which gives
 \es{knSimpAgain}{
  k_{n, s'} = \frac{64 \pi^3}{n^2 - s'^2} \int dr \, \frac{r^2}{\left(1 + r^2 \right)^3}\, \Z^{s', n}_{\mu_1\ldots \mu_s; \nu_1 \ldots \nu_s}(r \hat v) \,
    K^{\mu_1\ldots \mu_s; \nu_1 \ldots \nu_s}(r \hat v, 0) \,,
 }
where $\hat v$ is an arbitrary unit vector, say $\hat v = (0, 0, 1)$, and $\Z$ is a tensor ``zonal'' harmonic defined as
 \es{ZDef}{
  \Z^{s', n}_{\mu_1\ldots \mu_s; \nu_1 \ldots \nu_s}(x) \equiv \sum_{\ell, m}
   \HH_{\mu_1 \ldots \mu_s}^{s',n\ell m}(x)^* \HH_{\nu_1 \ldots \nu_s}^{s',n\ell m}(0) \,.
 }
We can thus find $k_{n, s'}$ by performing only a one-dimensional integral.  All that remains to do is to find explicit expressions for the tensor zonal harmonics $\Z^{s', n}$ and the kernel $K$.  We will do so in specific examples.

Before discussing the order $N^0$ corrections to $\delta F_\Delta^{(s)}$, however, we are already in position to show that the $\log N$ correction vanishes when $\Delta \neq s + 1$.  From~\eqref{knSimpAgain} and~\eqref{KS3}, we see that each $k_{n,s'}$ is proportional to $N$ and the normalization factor $C$.  The $\log N$ correction to $\delta F_\Delta^{(s)}$ is then found by evaluating the divergent sum
\es{deltaFVanish}{
\delta F_\Delta^{(s)} &= \left( \frac 12 \sum_{n=s+1}^\infty  \sum_{s'=-s}^s (n^2 - s'^2) \right) \log N + O(N^0) \\
&= \left( s+ {1 \over 2} \right) \left[ \zeta(-2, s+1) - {s(s+1) \over 3} \zeta(0,s+1) \right] \log N + O(N^0) \\
&= O(N^0)
}
through zeta function regularization.  In simplifying the second line above we have used a standard identity for the Hurwitz zeta-function.  We may use the same computation to show that (i) the $O(N^0)$ term does not depend on the normalization factor $C$, and (ii) if we reinstate the radius $R$ of the $S^3$, the potential $\log R$ term vanishes.  This latter point is important; since there is no anomaly in 3-d, the quantity $\delta F_\Delta^{(s)}$ must not have any dependence on the radius $R$ through terms that cannot be removed by the addition of local counter-terms.  A $\log R$ term is an example of such a term that cannot be removed.

\subsection{Particular cases}

We now calculate the order $N^0$ term in $\delta F_\Delta^{(s)}$ explicitly for $s = 0$, 1, and 2, and we show that the results are consistent with~\eqref{GenFormula}.  The $s = 0$ calculation has been performed in~\cite{Diaz:2007an,Klebanov:2011gs}, and as a warmup we begin by reviewing that computation.
We have also performed the $s=3$ calculation explicitly.  Some of the details may be found in Appendix~\ref{HARMONICSAPP} and~\ref{EVALUESAPP}.

\subsubsection{Spin $0$}

For $s = 0$ we have only one type of eigenvalue, $k_{n, 0}$.  Using~\eqref{KS3} we see that the kernel is given simply by
\es{KernelExplicit}{
  K(r \hat v, 0) =  \frac{N\,C}{\left(2 \sin (\chi/2) \right)^{2 \Delta}} \,,
 }
where we have defined
 \es{rDef}{
  r \equiv \tan \frac{\chi}{2} \,.
 }
 To compute the zonal harmonics we use the definition in~\eqref{ZDef} along with the explicit expressions for the spherical harmonics, given in appendix~\ref{spin0}, and we find
  \es{ZonalExplicit}{
  \Z^{0, n}(r \hat v) &= \sum_{\ell,m} Y_{n \ell m}(\chi, \theta, \phi) Y_{n \ell m} (\chi = 0) \\
  &= Y_{n 0 0}(\chi,\theta, \phi) Y_{n 0 0} (\chi = 0) = \frac{n \csc \chi \sin (n \chi)}{2 \pi^2} \,.
 }
The integral in \eqref{knSimpAgain} may then be performed explicitly:
 \es{knSpin0}{
  k_{n, 0} &= {N\, C\,2^{2(1- \Delta)\, }\pi \over n} \int_0^\pi d \chi \,{ \sin \chi \sin n \chi \over \left(\sin {\chi \over 2}\right)^{2 \Delta}} \\
  &= 4 \pi \, N \, C\sin (\pi \Delta) \frac{ \Gamma(2 - 2 \Delta) \Gamma(n-1 + \Delta )}{\Gamma(2 + n - \Delta)} \,.
 }

The change in the free energy may be evaluated using \eqref{deltaFGen}, which leads to the expression
\es{scalarSumSimp}{
\delta F_\Delta^{(0)} = {1 \over 2} \sum_{n = 1}^\infty n^2 \log {\Gamma(n-1 + \Delta ) \over \Gamma(2 + n - \Delta)} \,.
}
When $\Delta = {3 / 2}$ the operator $J^2$ is marginal, and so in that case we expect $\delta F_{3/2}^{(0)} = 0$.  Indeed, taking $\Delta = 3/2$ in~\eqref{scalarSumSimp}, we see that each of the terms in the sum vanishes independently.  The sum in~\eqref{scalarSumSimp} was evaluated explicitly for general $\Delta$ in~\cite{Diaz:2007an}, and their regularized result is
a particular case of \eqref{GenFormula}.  Below we give a more simple, though perhaps slightly less rigorous, derivation that will be useful when going on to the more complicated, higher-spin theories.  First we take a derivative of~\eqref{scalarSumSimp} with respect to $\Delta$, and then we insert a factor of $\exp[ - \epsilon \, n]$, $\epsilon > 0$, into the sum to make it convergent:
\es{scalarDerivation}{
\partial_\Delta \delta F_\Delta^{(0)} &={1 \over 2}  \frac{\partial^2}{\partial \epsilon^2} \left[  \sum_{n = 1}^\infty  \big[ \psi( 2 + n - \Delta) + \psi(n -1+\Delta) \big] e^{- \epsilon \, n} \right] \\
&= {3 - 2 \gamma - 2 \log \epsilon \over \epsilon^3} - {13 + 6 \Delta (\Delta - 3) \over 12\, \epsilon} \\
&+ {\pi \over 6} (\Delta - 1)\left( \Delta - {3 \over 2} \right) (\Delta - 2) \cot (\pi \Delta) + O(\epsilon) \,.
}
Subtracting the divergent terms from~\eqref{scalarDerivation} and using the relation $\delta F_\Delta^{(0)} =  \int^\Delta_{3/ 2} dx\, ( \partial_x \delta F_x^{(0)})$, which follows from the fact that $\delta F_{3/2}^{(0)} = 0$,  we arrive at the result in~\eqref{GenFormula} with $s = 0$.

\subsubsection{Spin $1$}

When $s=1$, a similar computation---using the results of Appendix~\ref{spin1} and~\ref{spin1Eigen}---gives
\es{results1}{
 k_{n, 0} &= N \, C {4 \pi (2 - \Delta ) \Gamma(2 - 2 \Delta) \sin ( \pi \Delta) \over \Delta} {\Gamma(n-1 + \Delta) \over \Gamma(n+2 - \Delta) }  \,, \\
  k_{n, \pm 1} &=   { \Delta - 1 \over 2 - \Delta}  k_{n, 0} \,.
}
This allows us to write $\delta F^{(1)}_\Delta$ as the sum
\es{deltaFvec2}{
\delta F^{(1)}_\Delta = {1 \over 2} \sum_{n = 2}^\infty \left[ n^2 \log k_{n, 0} + 2 (n^2 - 1) \log k_{n, 1} \right] \,,
}
which simplifies to
\es{deltaFvec3}{
\delta F^{(1)}_\Delta = {1 \over 2} \log \abs{ \Delta - 1 \over 2 - \Delta} + \sum_{n = 2}^\infty \left( {3 \over 2} n^2 - 1\right) \log \abs{ \Gamma(n - 1 + \Delta) \over \Gamma(n + 2 - \Delta) } \,.
}

As a first check of~\eqref{deltaFvec3}, we should verify that this expression vanishes when $\Delta = 3/2$.  Indeed, in this case each of the terms in the sum vanishes independently.  To evaluate~\eqref{deltaFvec3} for more general $\Delta$, it is again convenient to take a derivative with respect to $\Delta$ and to insert a factor of $e^{- \epsilon \, n}$ into the sum to make it convergent.  The identity
\es{ident1}{
\lim_{\epsilon \to 0^+}  \left[ {3 \over 2} \partial_\epsilon^2 - 1 \right] \sum_{n = 2}^\infty \left[ \psi(n + 2 - \Delta) + \psi(n -1 + \Delta) \right] e^{- \epsilon \, n} \\
 = {1 \over 2} \left( {1 \over 2 - 3 \Delta + \Delta^2} \right) + {\pi \over 2} \Delta (\Delta -3 ) \left(\Delta -{3 \over 2} \right) \cot (\pi \Delta)
}
then allows us to conclude that
\es{deltaFvec4}{
\partial_\Delta \delta F_\Delta^{(1)} = {\pi \over 2} \Delta (\Delta - 3) \left( \Delta - {3 \over 2} \right) \cot (\pi \Delta) \,,
}
which is consistent with~\eqref{GenFormula}.

\subsubsection{Spin $2$}

The calculation of the eigenvalues is again straightforward when $s = 2$, and it leads to
\es{svtTens}{
  k_{n, 0} &= c(\Delta) {\Gamma(n-1+\Delta) \over \Gamma(n+2 - \Delta)} \,, \qquad
   k_{n, 1} = {\Delta - 1 \over 2 - \Delta } k_{n, 0} \,, \\
   k_{n, 2} &=  {\Delta (\Delta - 1) \over (\Delta - 2) (\Delta - 3) } k_{n, 0} \,,
}
where the common factor
\es{cDelta}{
c(\Delta) = N \, C\, \frac{8 \pi(\Delta - 3)(\Delta - 2)(2 \Delta - 1) \Gamma(-2 \Delta) \sin (\pi \Delta)}{\Delta + 1}
}
is independent of $n$.
We then find that $\delta F^{(2)}_\Delta$ may be written as the sum
\es{deltaFtens2}{
\delta F^{(2)}_\Delta = \sum_{n=3}^\infty \left( {5 \over 2} n^2 - 5 \right) \log \abs{ \Gamma(n-1 + \Delta) \over \Gamma(n+2 - \Delta) } + {5 \over 2} \log \abs{ 2 \Delta^2 (\Delta - 1) \over (\Delta - 2) (\Delta - 3)^2 } \,.
}
When $\Delta = 3/2$, each of the terms in the sum vanishes identically, leading to the expected result $\delta F^{(2)}_{3/2} = 0$.
To evaluate this sum for more general $\Delta$, we follow the by now familiar procedure of taking a derivative with respect to $\Delta$ and inserting a factor $e^{- \epsilon \, n}$ into the sum to make it convergent.  Using an identity analogous to~\eqref{ident1}, we find  the result
\es{deltaFtens3}{
\partial_\Delta \delta F_\Delta^{(2)} = {5 \pi \over 6}(\Delta - 4) \left( \Delta - {3 \over 2} \right) (\Delta + 1) \cot (\pi \Delta) \,,
}
which is consistent with~\eqref{GenFormula}.

\subsection{A conjecture for arbitrary spin}

The spin 3 calculation is worked out explicitly in Appendix~\ref{SPIN3}.    From these examples with $s \leq 3$ we conjecture that at arbitrary integer spin $s$ the eigenvalues are related to each other by
\es{snvnArb}{
k_{n, 0} &= c_s(\Delta) {\Gamma(n-1+\Delta) \over \Gamma(n+2 - \Delta)} \,, \qquad k_{n, i} =  {\Gamma(2-\Delta) \over \Gamma(\Delta - 1) } { \Gamma(-1 + i + \Delta) \over \Gamma(2 + i - \Delta) } k_{n, 0}\,.
}
Importantly, the common factor $c_s(\Delta)$ is $n$-independent.  The calculation in~\eqref{deltaFVanish} that showed that $\delta F_\Delta^{(s)}$ does not depend on the radius $R$ and $N$ then also shows that $\delta F_\Delta^{(s)}$ is independent of $c_s(\Delta)$.  Moreover, when $\Delta = 3/2$ we find that $k_{n,i} = k_{n,0}$, which immediately implies that $\delta F_{3/2}^{(s)} = 0$.
To test the eigenvalue conjecture for more general $\Delta$, we may calculate $\partial_\Delta \delta F^{(s)}_\Delta$ using the identity
\es{regSumGen}{
&\lim_{\epsilon \to 0^+}  \left[ (1+2 s) \partial_\epsilon^2 - {s (1 + s)(1+2 s) \over 3} \right] \sum_{n = 1+s}^\infty \left[ \psi(n + 2 - \Delta) + \psi(n -1 + \Delta) \right] e^{- \epsilon \, n} \\
&+ {(1 +2 s) \over 3}\sum_{i = 1}^s \left( s(s+1) - 3 \, i^2 \right) \big[ \psi(2 - \Delta) + \psi(\Delta - 1) - \psi(2 + i - \Delta) - \psi(-1 + i + \Delta) \big] \\
&={(2 \, s + 1) \pi \over 3} \left( \Delta - {3 \over 2} \right) (\Delta + s - 1)(\Delta - s - 2) \cot (\pi \Delta) \,,
}
and we find the desired formula~\eqref{GenFormula} at arbitrary integer spin.  In section~\ref{spectral-zeta} we prove~\eqref{GenFormula} for arbitrary spin $s$ from a much simpler calculation in the bulk.

\section{Conserved Currents and Gauge Symmetries}
\label{conscurr}

Let us now return to the case where $\Delta = s +1$ and the operator $J_{\mu_1 \mu_2 \dots \mu_s}$ is a conserved current of spin $s$.
Specifically, we will consider the theories of $N$ free conformal complex scalars or Dirac fermion fields, which possess such currents of all $s>0$.
The conformal theory in this case is the gauge theory for the spin $s$ gauge field $h_{\mu_1 \mu_2 \dots \mu_s}$ with quadratic and higher-order
terms induced by the one-loop diagram with conformal matter propagating around the loop.  We will derive the result advertised in~\eqref{lNdivergence}, and we will also show explicitly that $\delta F$ is independent of the radius $R$ of the three-sphere. This independence of $R$ is crucial for the interpretation of the induced theory as a conformal theory.

For more generality, we work in $d$ dimensions, with $d$ odd.  The restriction to odd dimensions is put in to avoid the Weyl anomaly, which occurs when $d$ is even.  We return to the even dimensional case in later sections.  Note that in all $d$ the scaling dimension of the spin $s$ gauge field is $\Delta_- =2-s$.

The expression \eqref{lNdivergence}, as well as its generalization to arbitrary odd $d$, follows from a careful treatment of the gauge symmetry in the path integral. At the linearized level, the induced conformal higher-spin theory has the following local symmetries\footnote{We symmetrize with total weight one.  In other words $v_{(\mu_1 \mu_2 \dots \mu_s)}= \frac {1} {s!} \sum_{\sigma \in S_s} v_{\sigma_{\mu_1} \ldots \sigma_{\mu_s}}$.}
\begin{equation}
\delta h_{\mu_1\ldots\mu_s}=\nabla_{(\mu_1}v_{\mu_2\ldots \mu_s)}+g_{(\mu_1\mu_2}\lambda_{\mu_3\ldots\mu_s)} \,,
\label{conf-HS-gauge}
\end{equation}
where the rank $s-1$ symmetric traceless gauge parameter $v_{s-1}$ is the generalization of the familiar diffeomorphisms for spin 2, and the rank $s-2$ parameter $\lambda_{s-2}$ generalizes the local Weyl invariance of conformal gravity \cite{Fradkin:1985am}. We may use this symmetry to gauge away completely the trace of $h_{\mu_1\ldots\mu_s}$, and the remaining gauge symmetry is then obtained by restricting to the traceless part of (\ref{conf-HS-gauge})
\es{gaugeTrans}{
\delta h_{\mu_1 \mu_2 \dots \mu_s}=\big( {\cal O}_g v \big)_{\mu_1 \mu_2 \dots \mu_s}\,,
}
where the operator ${\cal O}_g$ takes the rank $s-1$ traceless symmetric tensor $v_{\mu_1 \ldots \mu_{s-1}}$ to a rank $s$ traceless symmetric tensor, namely:
\es{Og}{
\big( {\cal O}_g v \big)_{\mu_1 \mu_2 \dots \mu_s} =  \nabla_{(\mu_1} v_{\mu_2 \mu_3 \dots \mu_s)} - {s-1 \over d + 2(s - 2)} g_{(\mu_1 \mu_2 }\nabla^\nu v_{\mu_3 \mu_4 \dots \mu_{s}) \nu} \,.
}
One can then decompose the gauge field $h_{\mu_1 \dots \mu_s}$ as
  \es{hDecomp}{
   h_{\mu_1 \dots \mu_s} = t_{\mu_1 \dots \mu_s} + \big( {\cal O}_g v \big)_{\mu_1 \mu_2 \dots \mu_s} \,, \qquad
     \nabla^{\mu_1} t_{\mu_1 \dots \mu_s} = 0 \,.
  }
 The first term in \eqref{hDecomp} represents the physical modes, while the second term represents the pure gauge modes.  The requirement $\nabla^{\mu_1} t_{\mu_1 \dots \mu_s} = 0$ on the physical modes is a gauge fixing condition.

After integrating out the conformally invariant matter fields, the partition function at the conformal fixed point takes the form
\es{ZUV}{
Z &= \frac{1}{\Vol(G)} \int Dh e^{-S_\text{eff}[h]} \,,
}
where $G$ is the group of gauge transformations, and the effective action for the spin $s$ gauge field $h$ is given explicitly in the quadratic approximation by
\es{sphys}{
S_\text{eff}[h] = {1 \over 2} \int d^dx \sqrt{g(x)} \int d^dy \sqrt{g(y)} \, h^{i_1 \dots i_s}(x) K_{i_1 \dots i_s}\,^{j_1 \dots j_s}(x,y) h_{j_1 \dots j_s}(y) \,,
}
for some kernel $K$ as in~\eqref{KS3} for $d = 3$.   It is important that $K \propto N$, where $N$ is the number of conformally coupled matter fields;  when $N$ is large, the quadratic approximation \eqref{sphys} to the effective action becomes arbitrarily accurate.   The action $S_\text{eff}[h]$ is of course independent of the pure-gauge modes, so $S_\text{eff}[h] = S_\text{eff} [t]$.  Performing the split \eqref{hDecomp} and writing the volume of the group of gauge transformations as an integral over gauge parameters, we have
\es{ZUVAgain}{
Z &\approx {\int D({\cal O}_g v) \over \int Dv}\int Dt\,  e^{- S_\text{eff}[t]}  \,.
}
We are interested in studying the dependence on the $S^d$ radius $R$ and on the number $N$ of conformally coupled matter fields.  While only the last factor in \eqref{ZUVAgain} depends on $N$, the $R$-dependence of each of the two factors in \eqref{ZUVAgain} is more subtle. The absence of a Weyl anomaly guarantees, however, that $Z$ is independent of $R$, as we now explain.

On general grounds, the absence of a Weyl anomaly in odd dimensions means that the integration measure in the path integral is invariant under constant rescalings of the integration variables.  For instance, for a rank $s$ traceless symmetric tensor $h_{\mu_1 \ldots \mu_s}$, this means that $D h = D(\lambda h)$ for any constant $\lambda$.  We checked this fact explicitly in  \eqref{deltaFVanish} in $d=3$:  the Jacobian $D(\lambda h) / D h$ equals $\lambda$ raised to the sum of the degeneracies of all symmetric traceless tensor modes, and we checked that this sum vanishes in zeta-function regularization in $d=3$.  Similar checks are straightforward to perform for other odd $d$.

The action in \eqref{ZUV} remains unchanged if we send $g_{\mu\nu} \to \tilde \lambda^2 g_{\mu\nu}$ and $h_{i_1 \ldots i_s} \to \tilde \lambda^{s -2 + d/2} h_{i_1 \ldots i_s}$ (where $i_1, i_2, \ldots$ are frame indices).  Since the integration measure also remains unchanged (because all the modes are rescaled by the same factor), it follows that the partition function does not change either.   One then concludes that the partition function on $S^d$ is independent of $R$, because we can compute $Z$ for a sphere of unit radius, and then reinstate $R$ by performing a scale transformation.

In order to understand the dependence of \eqref{ZUVAgain} on $N$, we should first examine the zero modes of the operator ${\cal O}_g$.   These zero modes are important because in the numerator of the first factor in \eqref{ZUVAgain} we should not integrate over these modes, while in the denominator we should.   The zero modes of ${\cal O}_g$ are solutions to the conformal Killing tensor equation
\es{conformalKilling}{
\nabla_{(\mu_1} v_{\mu_2 \mu_3 \dots \mu_s)} = {s-1 \over d + 2(s - 2)} g_{(\mu_1 \mu_2 }\nabla^\nu v_{\mu_3 \mu_4 \dots \mu_{s}) \nu} \,.
}
 As shown in~\cite{1091.53020}, see also \cite{Lopatin:1987hz, Konstein:2000bi}, the symmetric traceless conformal Killing tensors of rank $s-1$ form an irreducible representation of $SO(d+1,1)$ of dimension
 \es{dimrep}{
n_{s-1} =  { (d  + 2 s -4) (d  + 2 s -3) (d  + 2 s -2) (d + s - 4)! (d+s - 3)! \over s! (s-1)! d! (d-2)! } \,.}
This is the representation of corresponding to the Young diagram
\es{young}{
\underbrace{\begin{picture}(200,60)
\put(0,10){\line(1,0){200}}
\put(0,30){\line(1,0){200}}
\put(0,50){\line(1,0){200}}
\put(0,10){\line(0,1){40}}
\put(20,10){\line(0,1){40}}
\put(40,10){\line(0,1){40}}
\put(60,10){\line(0,1){40}}
\put(80,10){\line(0,1){40}}
\put(140,10){\line(0,1){40}}
\put(160,10){\line(0,1){40}}
\put(180,10){\line(0,1){40}}
\put(200,10){\line(0,1){40}}
\put(110,20){\makebox(0,0){$\cdots$}}
\put(110,40){\makebox(0,0){$\cdots$}}
\put(210,14){\makebox(0,0)[l]{,}}
\end{picture}}_{\mbox{$s-1$}}
}
which has two rows of length $s-1$.\footnote{The same rectangular two-row representation appears naturally in the frame-like description of higher-spin gauge fields in $AdS_{d+1}$ \cite{Vasiliev:2001wa}.}  The representation may be labelled by the set of integers with $m_1=m_2=s-1$ and $m_3=\ldots=0$ corresponding to the length of each row, and we conventionally denote it as $[s-1,s-1]$.

Note that when $s = 2$,~\eqref{conformalKilling} reduces to the more familiar conformal Killing vector equation
\es{KillingVector}{
\nabla_\mu v_\nu + \nabla_\nu v_\mu  = {2 g_{\mu \nu} \over 3} \nabla \cdot v \,,
}
and it is well-known that there are $n_1 = (d+1)(d+2) / 2$ linearly independent conformal Killing vectors; they transform in the adjoint (antisymmetric two-index tensor) representation of $SO(d+1, 1)$.  An equivalent counting of conformal Killing tensors is in terms of representations of $SO(d+1)$, where the solutions of \eqref{conformalKilling} transform as irreps whose Young diagrams have two rows: $s-1$ boxes in the first row and any number of boxes in the second row.

We can now have a more detailed understanding of how each factor in \eqref{ZUVAgain} depends on $R$.  Let us start with the denominator of the first factor, $\Vol(G) = \int Dv$.  This quantity by itself is $R$-independent, as guaranteed by the absence of a Weyl anomaly and by the fact that we are integrating over all the modes of a rank $s-1$ traceless symmetric tensor.  We can split, however, the integral over all gauge parameters into an integral over the kernel of ${\cal O}_g$, which is the stabilizer of the gauge orbits, and an integral over the transverse space:
 \es{VolGSplit}{
  \Vol(G) = \Vol(H) \int D' v \,, \qquad \Vol(H) = \int_{\text{Ker} \, {\cal O}_g} Dv \,.
 }
The discussion above implies that $g_{\mu\nu} \propto R^2$, $t_{i_1 \ldots i_s} \propto R^{s-2 + d/2}$, and $v_{i_1 \ldots i_{s-1}} \propto R^{s-1 + d/2}$.  Since $\Vol(H)$ contains $n_{s-1}$ integrals and each integral contributes a factor of $R^{2-1 + d/2}$, we have
 \es{VolHRDependence}{
  \Vol(H) \propto R^{n_{s-1} (s - 1 + d/2)} \,, \qquad  \int D'v \propto R^{-n_{s-1} (s - 1 + d/2)} \,,
 }
where the $R$-dependence of $\int D'v$  is such that $\Vol(G)$ is $R$-independent.\footnote{The factor $\Vol(H)$ is also proportional to the volume of the gauge group.  While for $s=1$ the gauge group is compact, an extra complication that arises when $s>1$ is that the gauge group is now non-compact and its volume is formally infinite.}  The number of integration variables in $\int D'v$ is therefore equal to $-n_{s-1}$ in zeta-function regularization.  The $R$-dependence of the two other ingredients of \eqref{ZUVAgain} is
 \es{OtherRDependence}{
   \int D({\cal O}_g v) \propto R^{-n_{s-1} (s - 2 + d/2)} \,, \qquad \int Dt\,  e^{- S_\text{eff}[t]}  \propto R^{n_{s-1} (s - 2 + d/2)} \,.
 }
The first expression follows because that the number of integration variables equals $-n_{s-1}$ in zeta function regularization---for they're the same integration variables as in the $\int D'v$ integral---and because by dimensional analysis each integral contributes one fewer power of $R$ than each of the $\int D'v$ integrals.  The second expression in \eqref{OtherRDependence} is such that the $R$-dependence cancels when integrating over all rank-$s$ traceless symmetric tensor modes.  The number of integration variables equals $+n_{s-1}$ in zeta-function regularization, and each integral contributes a factor of $R^{s - 2 + d/2}$.

The dependence on $N$ in \eqref{ZUVAgain} comes entirely from the integrand of the second factor where $K \propto N$.  As a consequence of there not being a Weyl anomaly, we can write
\es{ZRescaled}{
  Z &\approx {\int D(\sqrt{N} {\cal O}_g v) \over \int D(\sqrt{N} v)}\int D(\sqrt{N} t)\,  e^{- S_\text{eff}[\sqrt{N} t]}  \,.
}
The second factor is now $N$-independent, while the first factor is proportional to $\left( 1/\sqrt{N}\right)^{n_{s-1}}$, simply because the denominator contains $n_{s-1}$ more integrals than the numerator.  Therefore
 \es{FlogNGEN}{
 \delta F = {n_{s-1} \over 2} \log N + O(N^0) \,,
 }
In $d = 3$, this expression reduces to~\eqref{lNdivergence}. This result was obtained in the leading large $N$ approximation where only the terms quadratic in the
 spin $s$ gauge field needed to be included in the induced action. In this approximation we could simultaneously gauge the currents with spins
 $s_1, s_2, \ldots, s_k$. In such a theory,
 \es{FlogNGEN2}{
 \delta F = \frac 1 2 \log N\sum_{i=1}^k n_{{s_i}-1} + O(N^0) \,.
 }
When non-linear effects are included in the induced gauge theory for higher-spin gauge fields, or equivalently in the dual Vasiliev theory in AdS$_{d+1}$ space, it may be necessary to gauge all the higher-spin symmetries simultaneously \cite{Vasiliev:2012vf}.

\subsection{The Chern-Simons terms}
\label{CS-bdy}

 In $d=3$, when the current is conserved and its dimension is $s+1$, we may add in a Chern-Simons term for the corresponding spin $s$ gauge field.\footnote{From the point of view of conformal invariance, this corresponds to the fact that in a 3-d CFT the 2-point function of a spin $s$ conserved current admits a conformally invariant parity odd contact term.}  It was shown in~\cite{Klebanov:2011td} that the Chern-Simons coefficient $k$ adds in quadrature with $N$:
\es{prinres}{
\delta F  =
\frac{1}{2} \log\left[ \pi \sqrt{ \left(\frac{N}{8}\right)^2 + \left( \frac{k}{\pi} \right)^2} \right]  \,.
 }
The special form of this answer follows from a formal $U(1)$ symmetry of the effective action for spin-$1$ gauge field coupled to
a conserved current.  The flat space action can be written as
\es{seff}{ S= \frac 12 \int {d^3 p\over (2\pi)^3} A^\mu (-p) A^\nu (p)
  K_{\mu\nu}(p) \ ,
 }
where, if we include the Chern-Simons term, the kernel $K_{\mu\nu}$ takes the form
\es{KmunuDefa}{
  K_{\mu\nu}(p) &= \frac{N}{16} \abs{p} \left(\delta_{\mu\nu} - \frac{p_\mu p_\nu}{\abs{p}^2} \right)
   +  \frac{k}{2 \pi} \epsilon_{\mu \nu}{}^{\rho} p_\rho \,.
 }
One can check that the effective action \eqref{seff} remains invariant under the infinitesimal transformation
\es{uonetrafo}
{\delta A_\nu (p) = \frac x 2 |p|^{-1}\epsilon_{\nu \alpha \beta} p^\alpha A^\beta(p) \,,
}
supplemented by the following transformation rules of the coefficients $k$ and $N$:
\es{trafores} { \delta \left ({k\over \pi}\right ) =- x {N \over 8}\ , \qquad \delta \left ({N\over 8}\right )= x {k \over \pi} \,.
}
Here, $x$ is an arbitrary small parameter.  The transformation \eqref{uonetrafo}--\eqref{trafores} does not commute with space-time parity because it mixes together the parity even and odd terms in the effective action.

The transformation rules \eqref{uonetrafo}--\eqref{trafores} can be exponentiated to obtain a $U(1)$ action on the modes $A_\mu(p)$ and the coefficients $k$ and $N$.  One obtains
 \es{FiniteA}{
   A_\mu(p ) &\to
  \left[   \left( 1 - \cos \frac x2\right)  \frac{p_\mu p_\nu}{ p^2}
   + \sin \frac x2  \frac{\epsilon_{\mu\nu\rho} p^\rho}{\abs{p}}
   + \cos \frac x2 \delta_{\mu\nu}  \right] A^\nu(p ) \,, \\
  \begin{pmatrix}
  k/\pi \\
  N/8
  \end{pmatrix} &\to
   \begin{pmatrix}
    \cos x & -\sin x \\
    \sin x & \cos x
   \end{pmatrix}
   \begin{pmatrix}
  k/\pi \\
  N/8
  \end{pmatrix} \,.
 }
The finite transformations \eqref{FiniteA} now leave the effective action \eqref{seff} invariant for any $x$;  they correspond to an $SO(2)$ symmetry under which the quantities $k/\pi$ and $N/8$ form a doublet.  This $SO(2)$ symmetry is not just a symmetry of the action, but it also leaves the integration measure invariant, because the $3 \times 3$ matrix appearing in the first line of \eqref{FiniteA} has unit determinant.  This symmetry explains why $\delta F$ depends only on the $SO(2)$ invariant combination $(N/8)^2 + (k/\pi)^2$.

This finding generalizes to $s>1$ where the action again has a parity even and a parity odd term.  For $s = 2$, the parity odd term is the well-known
 gravitational Chern-Simons term \cite{Deser:1982vy,Deser:1981wh}. The conformal gravity theory with only this term in the action was studied in \cite{Horne:1988jf}.
The effective action at quadratic order is \cite{Leigh:2003ez}
 \es{Seff2}{
  S =  \frac 12 \int {d^3 p\over (2\pi)^3} h^{\mu\nu} (-p) h^{\lambda \rho} (p)
  K_{\mu\nu, \lambda \rho }(p) \,,
 }
where the kernel $K$ can be written as the sum of an even-parity term with coefficient $C_T$ and an odd-parity term with coefficient $W_T$ in terms of the projector $\Pi_{\mu\nu}(p) = p^\mu p^\nu - \delta^{\mu\nu}p^2$:
 \es{Kspin2}{
  K_{\mu\nu, \lambda \rho }(p) &= C_T \frac{1}{2 \abs{p}} \left[\Pi_{\mu\lambda}(p) \Pi_{\nu\rho}(p)
   + \Pi_{\mu\rho}(p) \Pi_{\nu\lambda}(p)  - \Pi_{\mu\nu}(p) \Pi_{\lambda\rho}(p)\right]  \\
   &+ W_T \frac{p^\sigma}{4} \left[\epsilon_{\mu\lambda\sigma} \Pi_{\nu \rho}(p)
    + \epsilon_{\nu\lambda \sigma} \Pi_{\mu \rho}(p) + \epsilon_{\mu \rho \sigma} \Pi_{\nu \lambda}(p)
     + \epsilon_{\nu \rho \sigma} \Pi_{\mu \lambda}(p) \right] \,.
 }
One can check that this effective action is invariant under the infinitesimal transformations
 \es{InfinitesimalSpin2}{
   \delta h_{\mu\nu}(p) &= \frac{x}{4\abs{p}} \left( \epsilon_{\mu\lambda}{}^{\rho} p^\lambda h_{\rho \nu} + \epsilon_{\nu\lambda}{}^{\rho} p^\lambda h_{\mu \rho} \right)  \,, \\
   \delta C_T  &=  x W_T \,, \\
   \delta W_T &=  -x C_T \,,
 }
where $x$ is a small parameter.  Like in the $s=1$ case, these infinitesimal transformations exponentiate to finite $SO(2)$ transformations under which
 \es{FiniteSO2}{
  \begin{pmatrix}
   W_T \\
   C_T
  \end{pmatrix}
   \to   \begin{pmatrix}
    \cos x & -\sin x \\
    \sin x & \cos x
   \end{pmatrix}
   \begin{pmatrix}
   W_T \\
  C_T
  \end{pmatrix} \,.
 }
We expect $\delta F$ to depend only on the $SO(2)$-invariant $W_T^2 + C_T^2$:
 \es{deltaFSpin2}{
  \delta F = \frac 52 \log \left(W_T^2 + C_T^2 \right) \,.
 }

The discussion above should generalize to $s > 2$, where again the effective action is a sum of a parity-even term with coefficient $C$ and a parity-odd term with coefficient $W$.  The transformation rules are
 \es{TransfHigher}{
  \delta h_{\mu_1 \ldots \mu_s}(p) &= \frac{x}{2s}\epsilon_{(\mu_1 \nu}^{\rho} p^\nu h_{\rho \mu_2 \ldots \mu_s)}(p)
   \,, \\
  \delta C  &=  x W \,, \\
  \delta W &=  -x C \,.
 }
The change in the $S^3$ free energy due to the gauging of the spin $s$ current is then
 \es{deltaFs}{
 \delta F = \frac {(4 s^2-1) s}{12} \log \left(W^2 + C^2 \right)  \,.
}

Our result for $\delta F=F_\text{UV}-F_\text{IR}$ can be seen to be consistent with the structure of the $SL(2, \Z)$ action on the space of 3-d CFTs \cite{Witten:2003ya,Leigh:2003ez}.\footnote{For $s>1$ the $SL(2, \Z)$ is probably present only in the quadratic approximation to the induced action.
 We are grateful to E. Witten for discussions about this.} The $S$-generator of $SL(2, \Z)$ maps the theory with the ungauged spin $s$ current to the one with gauged higher-spin symmetry (the fixed point reached by the double-trace $J_s^2$ deformation), and vice-versa. Therefore, at the level of $\delta F$, the $S$ transformation essentially acts by exchanging $F_\text{UV}$ and $F_\text{IR}$, and therefore $\delta F$ should change sign under this operation. The $S$-generator transforms the parameters $W$, $C$ as
\begin{equation}
\tau \rightarrow -1/\tau \qquad \tau = W+i C
\end{equation}
or
\begin{equation}
C\rightarrow \frac{C}{W^2+C^2} \qquad W \rightarrow -\frac{W}{W^2+C^2}\,.
\end{equation}
It is then easy to see that, because of the logarithmic dependence on $W^2+C^2$, $\delta F$ indeed changes sign under this transformation.

\section{The calculation in AdS: general setup}

Let us consider a free massive spin $s$ field propagating in Euclidean AdS$_{d+1}$, i.e.~the hyperbolic space $\HH^{d+1}$. This can be described by a totally symmetric tensor\footnote{For $d=3$ a totally symmetric traceless tensor is the only possibility for a spin $s$ field. In higher dimensions, more general mixed symmetry fields are possible, but we will not consider them in this paper.} $h_{\mu_1 \cdots \mu_s}$ satisfying the Fierz-Pauli equations
\begin{eqnarray}
&&\left(\nabla^2-\kappa^2\right)h_{\mu_1 \cdots \mu_s}=0\,,\cr
&& \kappa^2=m^2-2+(s-2)(s+d-3)\,,\cr
&&\nabla^{\mu}h_{\mu\mu_2\cdots\mu_s}=0\,,\qquad g^{\mu\nu}h_{\mu\nu\mu_3\cdots\mu_s}=0\,.
\label{spin-s-EOM}
\end{eqnarray}
The mass term in the wave equation above is defined so that $m^2$ correspond to the physical mass of the field,\footnote{Except for $s=0$, where in this normalization $m=0$ gives a scalar with mass-squared equal to $4-2d$. For $d=3$, this is a conformally coupled scalar field.} while the extra spin-dependent shift arises from the coupling to the curvature of AdS (here and throughout we will set the AdS radius to one). These equations of motion and constraints may be derived from a Lagrangian, but we will not need the details of the general construction here. As a simple example, the $s=1$ case can be described by the Proca action
\es{Proca}{
S=\int d^{d+1}x \sqrt{g}\left(\frac{1}{4}F_{\mu\nu}F^{\mu\nu}+\frac{m^2}{2} A_{\mu}A^{\mu}\right)\,.
}
The equations of motion coming from this action, $\nabla^{\mu}F_{\mu\nu}=m^2A_{\nu}$, can be shown to be equivalent to (\ref{spin-s-EOM}) as long as $m^2\neq 0$. For massive fields, the equations (\ref{spin-s-EOM}) describe the propagation of $g(s)={ (2 s + d - 2) (s + d - 3)! \over (d-2)! s! }$ on-shell degrees of freedom.

In the massless case $m^2=0$, the spin $s\ge 1$ fields become gauge fields, with linearized gauge invariance
\es{gauge-s}{
\delta h_{\mu_1\cdots\mu_s}=\nabla_{(\mu_1}\epsilon_{\mu_2\cdots \mu_s)},
}
where the gauge parameter is a rank $s-1$ symmetric traceless tensor. The gauge invariant equations of motion and action are known \cite{Fronsdal:1978rb}, but we will not need their explicit form. The simple equations \eqref{spin-s-EOM} may be still used to describe the propagation of on-shell degrees of freedom. In this case, however, the second line of (\ref{spin-s-EOM}) does not follow from the equations of motion but can be imposed as a consistent on-shell gauge condition (see e.g. \cite{Mikhailov:2002bp}). Due to the usual counting of gauge symmetries, the number of propagating degrees of freedom in this case is
 \es{NumberOfDofs}{
g(s)-g(s-1)=\frac{(2 s + d - 3) (s + d - 4)! }{(d-3)! s! } \,.
 }
In $d+1=4$, this number gives $2$ degrees of freedom for all non-zero spins, corresponding to helicities $\pm s$.
In $d+1=3$ dimensions, there are no propagating degrees of freedom for $s>1$, and one for $s=1$.

The conformal dimension of the spin $s$ field theory operator dual to $h_{\mu_1\cdots\mu_s}$ can be obtained by studying the near-boundary behavior of a solution to the equations of motion.
To be concrete, if we use Poincar\'e coordinates for AdS$_{d+1}$
\es{Poincare}{
ds^2 = {dz^2 + \sum_{i = 1}^d dx_i^2 \over z^2} \,,
}
a solution to (\ref{spin-s-EOM}) behaves as $z\rightarrow 0$ as (see e.g. \cite{Giombi:2009wh})
$h_{i_1\cdots i_s} \sim z^{\Delta-s}$,
where $\Delta$ is a root of the equation
$(\Delta+s-2)(\Delta +2 - d - s)=m^2$.
The solutions to this equation are
\es{generalDim}{
\Delta_\pm = {d \over 2} \pm \nu \,, \qquad \nu = \sqrt{ m^2 + \left( {d \over 2} + s -2 \right)^2} \,.}
The same bulk theory describes two different CFTs depending on the boundary conditions for the field $h_{\mu_1 \cdots \mu_s}$, and these CFTs are exactly the endpoints of the RG flow obtained from the action in~\eqref{actionMod}. The boundary condition $h_{(s)} \sim z^{\Delta_- -s}$ corresponds to the UV CFT, with $J_s$ having dimension $\Delta_-$, and the boundary condition $h_{(s)} \sim z^{\Delta_+ -s}$ describes the IR fixed point, with $J_s$ of dimension $\Delta \equiv \Delta_+$. In the massless case, $\Delta_{+}=s+d-2$ is the dimension of the spin $s$ conserved current in the free theory, while $\Delta_{-}=2-s$ is the dimension of the spin $s$ auxiliary field that becomes a dynamical gauge field in the induced theory.

The contribution of $h$ to the free energy is given by evaluating the one-loop determinant
\es{contPhi}{
F^{(s)}_{\Delta_\pm} =\left. - \log  \int D h \,e^{ - s_h}  \right|_{\Delta_\pm} \,,
}
where the symbol $|_{\Delta_\pm}$ indicates which boundary conditions we are to impose at small $z$.  Thus, the change in free energy between the UV and IR fixed points is given by
\es{changeFPhi}{
\delta F^{(s)}_\Delta = F^{(s)}_{\Delta_-} - F^{(s)}_{\Delta_+} = {1 \over 2} \left[ \tr^{(s)}_- \log ( - \nabla^2 + \kappa^2) - \tr^{(s)}_+ 
\log ( - \nabla^2 + \kappa^2)  \right]  \,,
}
where the operator $\nabla^2=g^{\mu\nu}\nabla_\mu\nabla_\nu$ acts on symmetric transverse-traceless (STT) tensors of rank $s$.  Using the approach of
\cite{Gubser:2002zh,Diaz:2007an} and taking a derivative with respect to $\Delta$ gives the more convenient expression
\es{changeFPhi2}{
\partial_\Delta \delta F^{(s)}_\Delta = (2 \Delta - d) { \partial \delta F^{(s)}_\Delta\over \partial m^2} =  {2 \Delta - d \over 2}  \int \text{vol}_{\HH^{d+1}} \, \big( \text{Tr}\, G^{ (s) }_{\Delta_-} (x,x) - \text{Tr}\, G_{\Delta_+}^{ (s) } (x,x) \big)
}
in terms of the Green's functions $G^{ (s) }_{\Delta_\pm} (x,y)$ for the spin $s$ field with the respective boundary conditions. Here $\text{Tr}\, G^{(s)}(x,x)$ denotes the Green's function at coincident points traced over the space-time indices, namely $\text{Tr}\, G^{(s)}(x,x)=\lim_{y\rightarrow x} g^{\mu_1\nu_1}\cdots g^{\mu_s\nu_s}G_{\mu_1\ldots \mu_s\nu_1\ldots \nu_s}(x,y)$. Of course, the Green's function at coincident points is divergent, but the divergence is just the usual short-distance singularity of flat space propagators, which cancels when taking the difference between the two boundary conditions in \eqref{changeFPhi2} \cite{Gubser:2002zh}.

\section{Massive spin $s$ fields in AdS}

\subsection{Some lower spin examples}

As a warm-up we begin by considering a scalar field in $d = 3$.
In this case the Green's functions may be written down simply in terms of the chordal distance $u$.\footnote{The chordal distance $u$ is related to the geodesic distance $r$ by $u=\cosh r-1$.} Using the Poincar\'e coordinates \eqref{Poincare}, let us denote two points on AdS$_4$ by $x^{\mu}=(z,x^i)$ and $y^{\mu}=(w,y^i)$. Then the chordal distance is given by
\es{udeff}{
u(x,y) \equiv {(z - w)^2+(x^i-y^i)(x^i-y^i)  \over 2 z w} \,.
}
We then use the standard result for the Green's function of the massive scalar field on AdS$_{d+1}$ (see, for example, \cite{DHoker:1999ni}),
\es{scalarAdS}{
G_\Delta(x,y) &= G_\Delta(u) = \tilde C_\Delta (2 u^{-1} )^\Delta F\big(\Delta, \Delta - {d \over 2} + {1 \over 2} ; 2 \Delta - d + 1; - 2 u^{-1} \big) \,, \\
\tilde C_\Delta &= {\Gamma(\Delta) \Gamma(\Delta - {d \over 2} + {1 \over 2} ) \over (4 \pi)^{(d+1)/2} \Gamma(2 \Delta - d+ 1) } \,.
}
Taking $d = 3$, in the short-distance limit $u \to 0$ we find
\es{GsShort}{
G_\Delta(u) = {1 \over 8 \pi^2 u}  + O(\log u) \,,
}
and
\es{GsDiff}{
G_{3-\Delta} (u) - G_{\Delta} (u) = {1 \over 8 \pi} (\Delta - 1) (\Delta - 2) \cot ( \pi \Delta) + O(u) \,.
}
The only other ingredient needed to complete the computation is the regularized volume of $\HH^4$, which is $4 \pi^2 / 3$ (see \eqref{volHd}).  Combining this fact with~\eqref{GsDiff} and~\eqref{changeFPhi2} then allows us to reproduce~\eqref{GenFormula} with $s = 0$.

The spin $1$ calculation may be carried out in an analogous fashion to the spin $0$ calculation presented above.
The massive bulk-to-bulk vector field propagator was worked out explicitly in~\cite{Naqvi:1999va}:\footnote{The propagator we use, \eqref{vecProp}, differs by an overall minus sign compared to the one in \cite{Naqvi:1999va}. In these conventions the propagator reduces in the flat space limit to the Fourier transform of $(g_{\mu\nu}-k_{\mu}k_{\nu}/m^2)/(k^2+m^2)$.}
\es{vecProp}{
(G^{(1)}_{\mu \nu})_\Delta (u) =-\big[G_\Delta(u) + L_\Delta(u) \big] T_{\mu \nu} -{L'_\Delta(u)} S_{\mu \nu} \,,
}
where $G_\Delta(u)$ is the scalar propagator defined in~\eqref{scalarAdS} and
\es{LTS}{
L_\Delta(u) &= - {1 \over (\Delta - 1)(\Delta - 2) } \big[ 2 G_\Delta(u) + (1 + u) G_\Delta'(u) \big] \,, \\
T_{\mu \nu} &=  \partial_{\mu} \partial_\nu u \,, \qquad S_{\mu \nu} =  \partial_{\mu} u \partial_\nu u \,.
}
Using the explicit definition of $u$ in~\eqref{udeff}, we may work out that in the limit $u \to 0$ the trace $T_\mu^\mu \to - 4$ while $S_\mu^\mu \to 0$.  A straightforward calculation using the results above then leads to equation~\eqref{deltaFvec4}. One may perform an analogous computation using the massive spin $2$ propagator derived in \cite{Naqvi:1999va}. Following the same steps as above, one can evaluate the trace of the Green's function at coincident points.  Taking the difference of the two boundary conditions readily allows one to reproduce the CFT result (\ref{deltaFtens3}).

\subsection{Arbitrary spin}
\label{spectral-zeta}

In principle one may proceed to arbitrary spin by generalizing the method presented above for the spin $0$ and $1$ cases to general spin $s$.  Thankfully, however, there is a shortcut which saves us from having to solve for the massive bulk-to-bulk propagator at arbitrary spin.  Moreover, we may keep arbitrary the boundary spacetime dimension $d \geq 2$ in the following calculation without adding much complexity.  We begin by considering the integer spin cases, and we comment on the generalization to half-integer spin in Section \ref{half-int}.

Let us start by recalling the familiar definition of the heat kernel for the operator $-\nabla^2 + \kappa^2$ acting on transverse symmetric traceless spin $s$ tensors.  The heat kernel $K_{\mu_1 \cdots \mu_s}\,^{\nu_1 \cdots \nu_s}(x,x',t)$ on $\HH^{d+1}$ is a solution to the equations
\es{HeatKernel}{
\left({\partial \over \partial t}  -\nabla^2 + \kappa^2\right) K_{\mu_1 \cdots \mu_s}\,^{\nu_1 \cdots \nu_s}(x,y,t) &= 0 \,, \\
K_{\mu_1 \cdots \mu_s}\,^{\nu_1 \cdots \nu_s}(x,y,0) &= \delta_{(\mu_1 \cdots \mu_s)}\,^{(\nu_1 \cdots \nu_s)}(x,y) \,,
}
where $\delta_{(\mu_1 \cdots \mu_s)}\,^{(\nu_1 \cdots \nu_s)}(x,x')$ is the STT $\delta$-function on $\HH^{d+1}$. 
An explicit expression for the heat kernel may be written down in terms of the STT eigenfunctions $\hat h_{\mu_1 \cdots \mu_s}^{\lambda, u}$, which are taken to be orthonormal with respect to the standard inner product on $\HH^{d+1}$ and which satisfy the equation
\es{eigenfuncArb}{
-\nabla^2 \,\hat h^{\lambda, u}_{\mu_1 \cdots \mu_s}(x) = \left( \lambda^2 + {d^2 \over 4} + s \right) \hat h^{\lambda, u}_{\mu_1 \cdots \mu_s}(x)
}
as well as transversality and tracelessness.
Here $u$ is a multi-index labeling different eigenfunctions with the same eigenvalue under $-\nabla^2$, and it corresponds to the set of integers which specify the spherical harmonics on the $S^d$ boundary.  Additionally, the eigenvalue in (\ref{eigenfuncArb}) has been shifted in such a way that $\lambda \geq 0$.  In terms of these eigenfunctions, the heat kernel may be written formally as
\es{HeatKernel2}{
K_{\mu_1 \cdots \mu_s}\,^{\nu_1 \cdots \nu_s}(x,y,t) = \sum_u \int_0^\infty d \lambda\, &\hat h^{\lambda, u}_{\mu_1 \cdots \mu_s}(x) \hat h^{\lambda, u}\,^{\nu_1 \cdots \nu_s}(x')^* \\
&\exp\left[ - \left( \lambda^2 + {d^2 \over 4} + s + \kappa^2 \right) t \right] \,.
}
Note that using (\ref{spin-s-EOM}) and (\ref{generalDim}) we can write
\begin{equation}
\lambda^2 + {d^2 \over 4} + s + \kappa^2 =\lambda^2+\left(\Delta-\frac{d}{2}\right)^2\,,
\end{equation}
where $\Delta$ is the dimension of the dual operator.
The spectral zeta function $\zeta^{H}(z;x)$ is defined by evaluating the trace of the heat kernel at coincident points $x = y$, inserting a factor  of $t^{z-1}$, and integrating over $t$:
\es{spectralZeta}{
 \zeta^{H}(z;x) &\equiv {1 \over \Gamma(z) } \int_0^\infty dt \, t^{z-1} K_{\mu_1 \cdots \mu_s}\,^{\mu_1 \cdots \mu_s}(x,x,t) \\&=\sum_u \int_0^\infty d \lambda\, \frac{\hat h^{\lambda, u}_{\mu_1 \cdots \mu_s}(x) \hat h^{\lambda, u}\,^{\nu_1 \cdots \nu_s}(x)^*}{\left(\lambda^2+(\Delta- d/2)^2\right)^z}\,.
 }
Since the space $\HH^{d+1}$ is homogeneous, the zeta function does not depend on the position $x$.  We may define the integrated zeta function $\zeta^H(z)$ to be the integral of $\zeta^{H}(z,x)$ over the whole space, but for the reason just given this only has the effect of multiplying the expression in~\eqref{spectralZeta} by a factor of the regularized volume of $\HH^{d+1}$. This regularized volume may be found by writing the metric as $d\rho^2 + \sinh^2\rho\ d\Omega^2_{S^d}$ and imposing a cut-off on $\rho$ at a large value $\rho_c$. In even and odd dimensions this then gives \cite{Diaz:2007an, Casini:2010kt, Casini:2011kv}
 \es{volHd}{
 \int \text{vol}_{\HH^{d+1}} = \left\{ \begin{array}{ll}
 \pi^{d/2} \Gamma\left ( - {d \over 2} \right)  \,, & \text{$d$ odd} \,, \\
\frac{2\left(-\pi\right)^{d/2}}{\Gamma \left (1+\frac{d}{2} \right )}\log{ R }   \,, & \text{$d$ even} \,, \\
 \end{array} \right.
 }
 where $R$ is the radius of $S^d$ located at $\rho=\rho_c$.\footnote{Only the logarithmic divergence was retained in the even $d$ case. One may, for example, work in dimensional regularization with $d\rightarrow d-\varepsilon$, and identify the $1/\varepsilon$ pole with the $\log R$ divergence.} Since the integral over proper time $t$ of the heat kernel gives the Green's function, it is clear from the definition (\ref{spectralZeta}) that the spectral zeta function is related to the trace of the Green's function at coincident points by
 \es{relationG}{
 \zeta^H(z = 1) =  \int \text{vol}_{\HH^{d+1}}  \text{Tr}\, G_\Delta^{(s)} (x,x) \,.
 }
 The boundary conditions for the Green's function are determined by the boundary conditions we take for the eigenfunctions $h^{\lambda, u}_{\mu_1 \cdots \mu_s}(x)$.  The authors of~\cite{Camporesi:1993mz,Camporesi:1994ga} calculated $\zeta^H(z)$ for arbitrary spin and in arbitrary dimension $d$, assuming certain regularity conditions on the eigenfunctions that correspond to imposing the $\Delta_+$ boundary condition on the Green's function. To obtain the result for the $\Delta_-$ boundary condition, we will analytically continue their final result to arbitrary $\Delta$, as explained below.

Assuming for the moment $\Delta=\Delta_{+}$, the zeta function \eqref{spectralZeta}
may be written in terms of the integral over $\lambda$
 \es{interZeta}{
 \zeta^H(z) = \left({ \int \vol_{\HH^{d+1} } \over \int \vol_{S^{d}}} \right) {2^{d-1} \over \pi} g(s) \int_0^\infty d \lambda \, { \mu(\lambda) \over \left[ \lambda^2 + \left( \Delta_+ - {d \over 2} \right)^2 \right]^z } \,,
 }
with $g(s)$ the spin factor, which in $d = 2$ is given by $g(0) = 1$ and $g(s) = 2$ for $s \geq 1$, and in $d > 2$ by
\es{spinFactor}{
g(s) =
{ (2 s + d - 2) (s + d - 3)! \over (d-2)! s! }  \,, \qquad d \geq 3 \,.
}
This spin factor is the number of propagating degrees of freedom of a massive spin $s$ field in $d+1$ dimensions.
In $3+1$ dimensions, these are the familiar $2s+1$ degrees of freedom of a massive spin $s$ field.

The function $\mu(\lambda)$ is known as the spectral function, and it is obtained from \eqref{spectralZeta} by summing over all discrete indices of the eigenfunctions. The result of~\cite{Camporesi:1994ga} gives
\es{spectralMu}{
\mu(\lambda) = {\pi \left[ \lambda^2 + \left( s + {d - 2 \over 2} \right)^2 \right] \over \left( 2^{d-1} \Gamma \left (\frac {d+1} {2}\right ) \right)^2 } \abs{ \Gamma\left( i \lambda + {d - 2 \over 2} \right) \over \Gamma(i \lambda)}^2 \,.
}

We now turn to the evaluation of the integral in~\eqref{interZeta}, beginning with the case of most interest, $d = 3$.  The spectral function in $d = 3$ may be simplified to
\es{spectral3}{
\mu(\lambda) = {\pi \lambda \over 16} \left[ \lambda^2 + \left( s + {1 \over 2} \right)^2 \right] \tanh \pi \lambda \,,
}
and from this we see that to evaluate $\zeta^H(z)$ we need to compute the integral
\es{integral3}{
I_3(z)  = \int_0^\infty d \lambda \, \lambda \left[ \lambda^2 + \left( s + {1 \over 2} \right)^2 \right] { \tanh \pi \lambda \over \left[ \lambda^2 + \nu^2 \right]^z } \,, \qquad \nu \equiv  \Delta_{+} - {d \over 2}\,.
}
The integral only converges for Re$(z) > 2$, and so we proceed by assuming Re$(z) > 2$, evaluating $I_3(z)$ explicitly, and then analytically continuing to the other values of $z$.  One way to evaluate $I_3(z)$ is to use the identity $\tanh(\pi \lambda) = 1 - 2 (1 + e^{2 \pi \lambda} )^{-1}$ to write
\es{I32}{
I_3(z) =&{ \nu^{2 (1 - z) }\over 2 (2 - z)(1-z)} \left[ \nu^2 +(z-2) \left( s + {1 \over 2} \right)^2\right] \\
&-2\int_0^\infty d \lambda \, \lambda \left[ \lambda^2 + \left( s + {1 \over 2} \right)^2 \right] { 1\over (1 + e^{2 \pi \lambda}) \left[ \lambda^2 + \nu^2 \right]^z } \,.
}
The integral appearing above is now perfectly convergent for all $z$, and it may be evaluated explicitly for specific $z$ using, for example, the identities in~\cite{Camporesi:1991nw}.  The analytic continuation necessary to extract the result for $\Delta=\Delta_{-}$ can be done as follows. We first compute the integral (\ref{I32}) assuming $\Delta=\Delta_{+}$, so that $\nu\ge 0$. We then interpret the final result as an analytic function of $\nu$ (for instance, by replacing $|\nu|\rightarrow \nu$) and obtain the $\Delta_{-}=d-\Delta_{+}$ boundary condition by sending $\nu\rightarrow -\nu$.

An example of particular interest is $z = 1$, and in this case we find
\es{I3near1}{
I_3(z\approx 1) = &\left[ \left( s + {1 \over 2} \right)^2 - \nu^2 \right] {1 \over 2 (z - 1) } + \left[ \nu^2 - \left( s + {1 \over 2} \right)^2 \right] \psi\left( \nu + {1 \over 2} \right) \\
 &- {1 \over 24} - {\nu^2 \over 2} + O(z- 1) \,.
}

Substituting the result above into~\eqref{interZeta}, we obtain an expression for $\zeta^H(z \approx 1)$ with the $\Delta_+$ boundary condition. The pole at $z=1$ is just the expected short-distance singularity of the propagator, which will cancel when we compute the difference of the two boundary conditions $\zeta^H(z \approx 1) - \zeta^H_-(z \approx 1)$, where the minus subscript refers to the $\Delta_-$ boundary condition. As explained above, we find that a shortcut to obtaining $\zeta^H_-(z \approx 1)$  is to analytically continue the result in~\eqref{I3near1} letting $\nu \to - \nu$.\footnote{This analytic continuation becomes more subtle when $\Delta_+ = s + 1$, and so we treat this case separately later.}  Then, making use of the identity
\es{psiIdent}{
\psi\left( {1 \over 2} +\nu \right) - \psi\left( {1 \over 2} - \nu \right) = \pi \tan \nu \pi \,,
}
we obtain
\es{zetaDiff1}{
\left. \zeta^H(z ) -\zeta^H_-(z ) \right|_{z = 1} = - { \pi \over 3} \left( s + {1 \over 2} \right) (\Delta_+ - s -2)(\Delta_+ + s -1) \cot \pi \Delta_+ \,,
}
which, together with~\eqref{changeFPhi2}, immediately confirms the result for $\delta F^{(s)}_\Delta$ in~\eqref{GenFormula}.

The method used to derive~\eqref{zetaDiff1} becomes more cumbersome when generalizing to arbitrary space-time dimensions.  There is however a slightly more formal shortcut to evaluating~\eqref{interZeta} based on extending the region of integration in $\lambda$ to $(- \infty, + \infty)$ and closing the contour of integration in the complex plane.  One may then argue that
\es{zeta1GEN}{
\left. \zeta^H(z ) -\zeta^H_-(z ) \right|_{z = 1} = 2^d \left({ \int \vol_{\HH^{d+1} } \over \int \vol_{S^{d}}} \right)  g(s) {\mu\left[ i \left(\Delta_+ - {d \over 2} \right) \right] \over 2 \Delta_+ - d} \,.
}
When $d$ is odd we then find (even $d$ will be discussed in section \ref{Weylsec})
\es{finalOdd}{
\partial_\Delta \delta F^{(s)}_\Delta = (-1)^{(d-1)/2} g(s) {\Gamma\left(- {d \over 2} \right) \over 2^d \sqrt{\pi}\Gamma\big( \frac {d+1}{2}\big )}  &\left( \Delta - {d \over 2} \right) (\Delta + s -1) (\Delta - s -d +1) \\
&\Gamma(\Delta -1) \Gamma(d-1 - \Delta) \cos(\pi \Delta) \,.
}
Note that when $s = 0$ this agrees with the result in~\cite{Diaz:2007an,Klebanov:2011gs}. In $d=3$, it leads to the result quoted in eq.~(\ref{GenFormula}).

Moreover, we conjecture the identity
\es{zeta0GEN}{
\left. \zeta^H(z ) -\zeta^H_-(z ) \right|_{z = 0} = 2^{d} \left({ \int \vol_{\HH^{d+1} } \over \int \vol_{S^{d}}} \right)  g(s)\,  i \left(\res_{\lambda = i (\Delta_+ - d/2)}   \mu(\lambda) \right) \,,
}
which is useful when $\Delta_+ = d + s - 2$, corresponding to a conserved current at the boundary.  Note that this expression vanishes in even $d$ for all $\Delta_+$ and vanishes in odd $d$ when $\Delta_+ \neq d + s - 2$.  However, when $\Delta_+ = d + s - 2$ and $d$ is odd, we find
\es{zeta0GEN2}{
\left. \zeta^H(z ) -\zeta^H_-(z ) \right|_{z = 0} = - n_{s-1} \,,
}
with $n_{s-1}$ defined in~\eqref{dimrep}.  We will explain the significance of these results in the next section.

\section{Massless higher-spin fields in AdS and gauge symmetries}
\label{masslessAdS}

In this section we discuss directly the case of massless higher-spin fields, the corresponding gauge fixing and the bulk interpretation of the coefficient of $\log N$ associated to the $\Delta_{-}$ boundary conditions.
As usual, in computing the one-loop partition function for a higher-spin gauge field, we must properly gauge fix the local symmetry (\ref{gauge-s}). Using a covariant gauge fixing procedure and introducing the corresponding ghosts,\footnote{Alternatively, one may use a procedure similar to the one discussed in Section 4 by explicitly decomposing the higher-spin gauge field into its transverse, trace and pure gauge parts.} the end result is that the one-loop partition function in AdS$_{d+1}$ may be written as the ratio of determinants (see for example \cite{Gibbons:1978ac, Gibbons:1978ji, Christensen:1979iy, Yasuda:1983hk} for the spin 2 case, and \cite{Gaberdiel:2010ar, Gaberdiel:2010xv, Gupta:2012he} for the generalization to arbitrary spin)
\es{one-loop-gauge}{
Z_{(s)}=\frac{\left[{\rm det}^{STT}_{s-1}\left(-\nabla^2+(s-1)(d+s-2)\right)\right]^{\frac{1}{2}}}{\left[{\rm det}^{STT}_{s}\left(-\nabla^2+(s-2)(d+s-3)-2\right)\right]^{\frac{1}{2}}} \,,
}
where each determinant is computed on the space of symmetric traceless transverse tensors. The numerator corresponds essentially to the spin $s-1$ ghost contribution. The structure of the associated kinetic operator may be obtained basically by ``squaring" the gauge transformation
\begin{equation}
\begin{aligned}
&\int d^{d+1}x \sqrt{g} \nabla_{(\mu_1}\xi_{\mu_2\ldots \mu_s)}\nabla^{(\mu_1}\xi^{\mu_2\ldots \mu_s)}\\
&=\int d^{d+1}x \sqrt{g} \xi^{\mu_1\ldots \mu_{s-1}} \left(-\nabla^2+(s-1)(d+s-2)\right) \xi_{\mu_1\ldots \mu_{s-1}} \,,
\end{aligned}
\label{ghost-kin}
\end{equation}
where we have integrated by parts, restricted to transverse $\xi_{s-1}$, and related commutators of covariant derivatives to the curvature of AdS (we set the AdS radius to one).

Recall that we are interested in computing the ratio of the partition functions with $\Delta_{+}=d+s-2$ and $\Delta_{-}=2-s$ boundary conditions imposed on the physical spin $s$ gauge fields. However, when computing the ghost determinant in (\ref{one-loop-gauge}), we also have in principle two choices of boundary behavior for the Green's function associated to the kinetic operator $-\nabla^2+(s-1)(d+s-2)$. Working in Poincare coordinates and using (\ref{generalDim}), one finds that the two boundary conditions on the spin $s-1$ transverse field with such kinetic operator are
\begin{equation}
\xi_{i_1\ldots i_{s-1}}(z,x_i) \sim z^{\delta_{\pm}}c_{i_1\ldots i_{s-1}}(x_i),\qquad \delta_{+}=d,\qquad \delta_{-}=2-2s \,,
\label{ghost-bc}
\end{equation}
where $i_1,\ldots,i_{s-1}$ are indices along the flat $d$-dimensional boundary. As we now explain, the choice of $\delta_{\pm}$ ghost behavior is correlated with the choice $\Delta_{\pm}$ on the physical gauge field. To see this, we can look at the structure of the allowed gauge transformations on the spin $s$ gauge field
\begin{equation}
\delta h_{\mu_1\ldots\mu_s}=\nabla_{(\mu_1}\xi_{\mu_2\ldots\mu_s)}\,.
\label{gauge-tr}
\end{equation}
The boundary behavior of the gauge field is
\begin{equation}
h_{i_1\ldots i_s}(z,x_i) \sim z^{\Delta_{\pm}-s} \alpha_{i_1\ldots i_s}(x_i)\,,\qquad \Delta_{+}=s+d-2\,,\qquad \Delta_{-}=2-s\,.
\label{boundary-h}
\end{equation}
In the case of the ordinary $\Delta_{+}$ boundary condition, we see that in order for the gauge transformation to preserve the boundary behavior of
the spin $s$ gauge field, we must choose in \eqref{ghost-bc} the $\xi_{s-1} \sim z^{d}$ behavior for the ghost.  The bulk gauge transformations then fall off fast enough at the boundary so that the bulk spin $s$ field is dual to a gauge invariant conserved current.
On the other hand, with the alternate $\Delta_{-}$ boundary condition, $h_{(s)}$ is dual to a gauge field at the boundary. In this case, we expect that the bulk gauge transformations should reproduce in the $z\rightarrow 0$ limit the gauge transformations in the boundary theory. From (\ref{ghost-bc}), we see that the $\delta_{-}=2-2s$ behavior for the ghost is precisely what we need for this to happen, since in this case the spin $s$ gauge field (\ref{boundary-h}) and the ghost have the same scaling in the boundary limit.

In section \ref{conscurr} we explained that the coefficient of $\log N$ in the free energy can be understood as counting the numbers of missing gauge transformations, or equivalently ghost zero modes. We thus expect that an analogous interpretation should hold in the bulk. Indeed, the quadratic action for the bulk spin $s$ fields has the schematic form
\begin{equation}
S\sim N \int d^{d+1}x \sqrt{g} h_{(s)} {\cal D}^{(s)} h_{(s)} \,,
\end{equation}
where $N$ plays the role of the (inverse of the) coupling constant. The ghost action does not carry $N$ dependence. However, by general arguments (see e.g. \cite{Witten:1995gf} for a related discussion), the Gaussian path integral on the spin $s$ field gives a coupling dependence in the partition function
\begin{equation}
\left(\frac{1}{\sqrt{N}}\right)^{d_s-(d_{s-1}-n_{s-1})} \,,
\label{N-scaling-Z}
\end{equation}
where $d_s$ is the dimension of the space of unconstrained spin $s$ fields, $d_{s-1}$ the dimension of the spin $(s-1)$ gauge parameter space, and $n_{s-1}$ the number of gauge transformations that act trivially on the gauge field. Using a regularization such that $d_s=d_{s-1}=0$ (such as the $\zeta$-function regularization we used in the boundary), the $N$ dependence of the one-loop free energy will then be $F=\frac{1}{2}n_{s-1}\log N$. To prove agreement with the boundary calculation, we just have to show that we have the same number $n_{s-1}$ of trivial gauge transformations (or ghost zero modes) in the bulk as we do in the boundary, and also, importantly, that such zero modes of the gauge transformation are only present with the $\Delta_{-}$ boundary condition.

The trivial bulk gauge transformations that we should count are the solutions to
\begin{equation}
\nabla_{(\mu_1}\xi_{\mu_2\ldots\mu_s)}=0,\qquad \xi^{\mu}_{\ \mu\mu_3\ldots\mu_{s-1}}=0\,;
\label{Killing-tensor}
\end{equation}
namely, they are the traceless spin $s-1$ Killing tensors of the AdS background. Note that due to (\ref{ghost-kin}) these are also zero modes of the ghost kinetic operator. The traceless Killing tensors of AdS$_{d+1}$ are expected to be in one-to-one correspondence with the conformal Killing tensors in the boundary CFT \cite{Mikhailov:2002bp}. So we anticipate that solutions to (\ref{Killing-tensor}) should fall into the $[s-1,s-1]$ representation of $SO(d+1,1)$, and hence we should have the same number of zero modes in the bulk and in the boundary. However, since the boundary behavior of these modes is crucial in our analysis, it is important to analyze explicitly the solutions to (\ref{Killing-tensor}).

Let us first look at the simplest $s=1$ case. Here we are just counting solutions to
\begin{equation}
\nabla_{\mu}\xi = 0\,.
\label{spin0-ghost}
\end{equation}
Clearly the only solution is $\xi = {\rm constant}$ over the whole AdS. If the gauge field is quantized with the $\Delta_{+}$ boundary condition, then, as we have argued above, the analysis of allowed gauge transformations requires $\xi \sim z^{d}$ near the boundary. Therefore, as expected, this constant mode should not be counted as a trivial gauge transformation in the $\Delta_{+}$ theory. On the other hand, with the $\Delta_{-}$ boundary condition the scalar ghost should have precisely the behavior $\xi \sim z^0$ at small $z$ (see (\ref{ghost-bc})), and so the constant mode solving (\ref{spin0-ghost}) should indeed be intepreted as a trivial gauge transformation of the $\Delta_{-}$ theory. Of course, the projection of this mode to the boundary (trivially) coincides with the single constant gauge transformation on $S^3$, leading to $\delta F_{s=1}=1/2\log N+O(N^0)$.

For $s=2$, we should look for solutions to
\begin{equation}
\nabla_{\mu}\xi_{\nu}+\nabla_{\nu}\xi_{\mu}=0\,.
\end{equation}
These are just the Killing vectors generating the isometries of AdS$_{d+1}$, and the solution is well known. There are $(d+1)(d+2) / 2$ Killing vectors transforming in the adjoint representation of $SO(d+1,1)$. We may describe (Euclidean) AdS$_{d+1}$ as the hyperboloid in $\mathbb{R}^{d+1,1}$
\begin{equation}
\eta_{AB}X^A X^B=-1 \qquad A,B=0,1,\ldots d+1\,,
\end{equation}
where $\eta_{AB}=(-1,+1,\ldots,+1)$. Choosing an explicit parameterization $X^A(x^{\mu})$, where $x^{\mu}$ are coordinates on AdS$_{d+1}$, the Killing vectors are given by
\begin{equation}
\xi^{AB}_{\mu}=X^A \partial_{\mu} X^B-X^B\partial_{\mu}X^A\,.
\end{equation}
For instance, in the Poincare coordinates
\begin{equation}
X^A=\left(\frac{z}{2}\big[1+\frac{1}{z^2}(1+z^2+x^i x^i)\big],\frac{x^i}{z},\frac{z}{2}\big[1+\frac{1}{z^{2}}(1-z^2-x^i x^i) \big]\right),\qquad i=1,\ldots,d \,.
\end{equation}
A simple calculation shows that the Killing vectors behave at small $z$ as
\begin{equation}
\xi_i^{AB} = z^{-2} v_i^{AB}(x_i)+O(z^0), \qquad \xi_z^{AB} = z^{-1} f^{AB}(x_i)\,.
\end{equation}
From (\ref{ghost-bc}) and the discussion thereafter, we conclude that these are truly zero modes of the bulk gauge transformations only when the graviton is quantized with the alternate $\Delta_{-}$ boundary condition. Therefore, we reproduce the result $F_{\Delta_{-}}^{(2)}-F_{\Delta_{+}}^{(2)}=5 \log N$ in $d=3$. As a remark, note that the boundary limit of the AdS Killing vectors yields as expected the conformal Killing vectors on the boundary, as one can explicitly check\footnote{We have used Poincar\'e coordinates for simplicity in discussing the boundary behavior. However this result is general. For instance, using the metric
$d\rho^2 + \sinh^2\rho\ d\Omega^2_{S^d}$ one can reproduce the conformal Killing vectors on $S^d$ from the $\rho\rightarrow \infty$ limit of the AdS Killing vectors.}
\begin{equation}
\lim_{z\rightarrow 0} z^2 \xi_i^{AB} = v_i^{AB}(x_i),\qquad
\nabla_iv_j^{AB}+\nabla_j v_i^{AB}-\frac{2}{d}g_{ij}\nabla^kv_k^{AB} = 0\,.
\end{equation}

To proceed with the higher-spin cases, we can use the result \cite{Thompson:1986tt} that in spaces of constant curvature (such as AdS) all Killing tensors of rank greater than or equal to two are reducible; i.e.~they can be constructed from symmetrized tensor products of the Killing vectors. It is clear that when we take the tensor product of $s-1$ Killing vectors, which transform in the $[1,1]$ representation, we get a sum of irreducible representations including in particular $[s-1,s-1]$. In fact, after imposing that the resulting Killing tensor is traceless in the spacetime indices, all representations except $[s-1,s-1]$ are projected out. Let us see this more explicitly. At rank $s-1$, we construct the symmetric tensor
\begin{equation}
\xi_{\mu_1\ldots \mu_{s-1}} = C_{A_1B_1,A_2B_2,\ldots,A_{s-1}B_{s-1}} \left[\xi^{A_1B_1}_{\mu_1}\xi^{A_2B_2}_{\mu_2}\cdots \xi^{A_{s-1}B_{s-1}}_{\mu_s}+\ldots\right] \,,
\label{K-tensor-sol}
\end{equation}
where the term in the square brackets is completely symmetrized in the spacetime indices, and $C_{A_1B_1,\ldots,A_{s-1}B_{s-1}}$ is a constant tensor, which, by construction, is antisymmetric in each pair of indices and symmetric under exchange of any pair. It is easy to see that this solves the Killing tensor equation, and the theorem guarantees that there are no additional non-trivial solutions in AdS. To impose the tracelessness condition, we note that the Killing vectors satisfy an indentity of the form
\begin{equation}
g^{\mu\nu}\xi_{\mu}^{AB}\xi_{\nu}^{CD}=
\frac{1}{d}\left[\eta^{AC}\xi^{EB}_{\mu}\xi^{\mu \ D}_E\pm {\rm 3~terms} -\frac{1}{d-1}(\eta^{AC}\eta^{BD}-\eta^{AD}\eta^{BC})\xi^{EB}_{\mu}\xi^{\mu}_{EB}\right]\,.
\end{equation}
Therefore, as long as all traces are removed from the coefficient tensor $C_{A_1B_1,\ldots,A_{s-1}B_{s-1}}$, we obtain a traceless Killing tensor. Finally, we note that if $C_{A_1B_1,\ldots,A_{s-1}B_{s-1}}$ were totally antisymmetric in 3 or more indices, (\ref{K-tensor-sol}) would vanish identically. To summarize, $C_{A_1B_1,\ldots,A_{s-1}B_{s-1}}$ is constrained to be antisymmetric in each pair of indices, completely traceless, and such that the antisymmetrization over any 3 indices gives zero. Indeed, this can be seen to be a realization of the $[s-1,s-1]$ representation of $SO(d+1,1)$. As a familiar example, at $s=3$ we see that $C_{A_1B_1,A_2B_2}$ is constrained to have the symmetries of the Weyl tensor (in $d+2$ dimensions), which correspond to the $[2,2]$ representation. From the explicit tensor product construction, it is clear that the boundary behavior of these traceless Killing tensors is $\xi_{i_1\ldots i_{s-1}}\sim z^{2-2s}$. From (\ref{ghost-bc}), we see that this is precisely the behavior we should impose on the ghosts when the spin $s$ field is quantized with the $\Delta_{-}$ boundary condition. Therefore, we find the expected $n_{s-1}={\rm dim}([s-1,s-1])$ ``missing" gauge transformation in the $\Delta_{-}$ theory and reproduce from the bulk the result
\begin{equation}
F^{(s)}_{\Delta_{-}}-F^{(s)}_{\Delta_{+}} = \frac{1}{2}n_{s-1}\log N+O(N^0)\,.
\label{deltaF-logN}
\end{equation}

To conclude this section, let us observe that it appears to be possible to reproduce the correct coefficient of $\log N$ also by some formal manipulations on the spectral $\zeta$-function, as discussed in Section \ref{spectral-zeta}. Because the overall coupling in front of the bulk higher-spin action is proportional to $N$, the coefficient of $\frac{1}{2}\log N$ can be understood (see (\ref{N-scaling-Z})) as counting the dimension of the space of the physical spin $s$ field. Therefore, we may try to formally compute
\es{logN}{
\delta F_\Delta^{(s)} = {\log N \over 2} \left( \tr_-^{(s)}  - \tr_+^{(s)} \right) = -\left. { \log N \over 2} \left[  \zeta^H(z ) -\zeta^H_-(z )  \right] \right|_{z = 0} \,.
}
From the discussion of the previous section we see that this expression vanishes unless  $\Delta = d+ s -2$ and $d$ is odd.  In that case, we may use the result in~\eqref{zeta0GEN2} to calculate $\delta F_{d+s-2}^{(s)}$, and one can see that this indeed leads to the expected result.

\subsection{Mixed boundary conditions and Chern-Simons terms}
In the previous section we concentrated on the case of the two $\Delta_{\pm}$ boundary conditions. In fact, for gauge fields in AdS$_4$, a more general mixed boundary condition is possible. In this section, we restrict to $d=3$ and make some comments on these mixed boundary conditions and their relation to boundary Chern-Simons terms.

Let us examine more closely the boundary conditions for the massless spin 1 field in $d=3$. The components of the gauge field in AdS$_4$ solving the equations of motion (\ref{spin-s-EOM}) have the following small $z$ behavior,\footnote{The small $z$ expansion of $A_z$ can be related to the one of $A_i$ by the gauge condition $\nabla_{\mu}A^{\mu}=0$.}
\es{spin1-bc}{
A_i(z,\vec{x})=\alpha_i(\vec{x})+z \beta_i(\vec{x})+O(z^2)\,,\qquad A_z(z,\vec{x}) = O(z) \,.
}
The regular $\Delta_{+}$ boundary conditions correspond to $\alpha_i=0$, and then $\beta_i(\vec{x})$ is dual to the conserved spin 1 current in the boundary. The alternate $\Delta_{-}$ boundary conditions correspond to $\beta_i=0$, and $\alpha_i(\vec{x})$ is dual to the dynamical gauge field at the boundary. Equivalently, these boundary conditions can be expressed in a gauge invariant form respectively as vanishing of the boundary magnetic field
\es{electric}{
F_{ij}|_{z=0} =0
}
or vanishing of the boundary electric field
\es{magnetic}{
F_{zi}|_{z=0}=0\,.
}
More generally, one may impose a one parameter family of conformally invariant boundary conditions \cite{Witten:2003ya} (see also \cite{Chang:2012kt} for a detailed discussion)
\es{mixed}{
\frac{1}{2}\epsilon_{ijk}F_{jk}+ib_1\frac{N}{k_1}F_{zi}\Big{|}_{z=0}=0\,,
}
where $b_1$ is a constant which depends on the normalization of the boundary 2-point function of the spin 1 current (in our conventions, $b_1= \pi /8$). For finite $k_1$, these boundary conditions correspond to gauging the $U(1)$ global symmetry at the boundary, while adding a Chern-Simons term at level $k_1$. Indeed, note that in terms of the expansion in (\ref{spin1-bc}), these boundary conditions amount to $k_1\epsilon_{ijk}\partial_j \alpha_k + i b_1 N \beta_i=0$, which is the structure of the equations of motion for the boundary gauge field in the presence of the Chern-Simons term, $\beta_i$ playing the role of the current. The regular $\Delta_{+}$ boundary conditions are recovered in the limit $k_1\rightarrow \infty$, while $k_1=0$ gives the $\Delta_{-}$ boundary condition dual to conformal QED with no Chern-Simons term. The Green's function for the spin 1 field with these boundary conditions was worked out (in the $A_z=0$ gauge) in \cite{Chang:2012kt}, and one can explicitly see that in the $z\rightarrow 0$ limit it reproduces the 2-point function of the boundary gauge field with Chern-Simons term, namely the inverse (in the gauge $\partial_i A^i=0$) of the kinetic operator (\ref{KmunuDefa}). Introducing the self-dual and anti self-dual parts of the field strength $F_{\mu\nu}^{\pm}=F_{\mu\nu}\pm\frac{1}{2}\epsilon_{\mu\nu}^{\ \ \rho\sigma}F_{\rho\sigma}$, the mixed boundary conditions may be also written as
\begin{equation}
e^{i\gamma_1} F^{+}_{zi}\Big{|}_{z=0}=e^{-i\gamma_1}F^{-}_{zi}\Big{|}_{z=0}\,,\qquad \quad
e^{i\gamma_1} = \sqrt{\frac{k_1+i b_1 N}{k_1-i b_1 N}}\,.
\label{spin1-Fpm}
\end{equation}
In this form, the ordinary and alternate boundary conditions correspond respectively to $\gamma_1=0$ and $\gamma_1 = \frac{\pi}{2}$.

The possibility of imposing conformally invariant mixed boundary conditions extends to the higher-spin cases. From the boundary point of view, it corresponds to the fact that we can add parity-odd local conformal actions of Chern-Simons type for the higher-spin gauge fields \cite{Fradkin:1989xt}. In the spin 2 case, this is just the familiar gravitational Chern-Simons action $i k_2 \int {\rm tr} \left(\omega \wedge d\omega+\frac{2}{3} \omega^3\right)$. The mixed boundary conditions for spin 2 were discussed for instance in \cite{Compere:2008us, deHaro:2007fg, deHaro:2008gp}. A solution to the linearized $s=2$ equations of motion (\ref{spin-s-EOM}) has the small $z$ behavior
\begin{equation}
h_{i_1 i_2}(z,\vec{x})=\frac{1}{z^2} \alpha_{i_1i_2}(\vec{x}) +\ldots + z \beta_{i_1 i_2}(\vec{x})+O(z^2)\,.
\label{spin2-bc}
\end{equation}
To express the boundary conditions in a gauge covariant form similar to the spin 1 treatment above, we note that the natural generalization of the spin 1 field strength is the Weyl tensor $C_{\mu\nu\rho\sigma}.$\footnote{We use conventions in which the Weyl tensor satisfies $C_{\mu\nu\rho\sigma}=-C_{\nu\mu\rho\sigma}=-C_{\mu\nu\sigma\rho}$, $C_{\mu\nu\rho\sigma}=C_{\rho\sigma\mu\nu}$, $C_{[\mu\nu\rho]\sigma}=0$, and it is completely traceless.} On a solution to the equations of motion, one finds that $\alpha_{i_1i_2}(\vec{x})$ and $\beta_{i_1i_2}(\vec{x})$ in (\ref{spin2-bc}) are related to the ``electric" and ``magnetic" components of the Weyl tensor \cite{Compere:2008us}\footnote{To derive this result, one can solve the equations of motion and gauge conditions in (\ref{spin2-bc}) perturbatively in small $z$. The terms which are omitted in the expansion (\ref{spin2-bc}) are determined in terms of $\alpha_{i_1i_2}$ by the equations of motion.}
 \es{Weyl-small-z}{
z C_{z i z j}\Big{|}_{z=0} &=-\frac{3}{2}\beta_{i j}(\vec{x})\,,\qquad \\
z \frac{1}{2}\epsilon_{ikl} C_{kl z j}\Big{|}_{z=0}&= \frac{1}{8}\left[\left(\epsilon_{ik m}\partial_m \left(\delta_{jl}\partial^2-\partial_j\partial_l\right)+(i \leftrightarrow j)\right)+(k \leftrightarrow l)\right]\alpha_{kl}(\vec{x})\,.
}
Therefore we see that the regular $\Delta_{+}$ and alternate $\Delta_{-}$ boundary conditions may be expressed respectively as vanishing of the ``magnetic" part of the Weyl tensor
\begin{equation}
\frac{1}{2}\epsilon_{ikl} C_{kl z j}\Big{|}_{z=0}=0
\end{equation}
or vanishing of the ``electric" part
\begin{equation}
C_{z i z j}\Big{|}_{z=0}=0\,,
\end{equation}
in complete analogy with (\ref{electric}) and (\ref{magnetic}).
Note that if $\alpha_{ij}$ is viewed as a linearized perturbation of the boundary flat metric, then one can see that the three-derivative operator in the second line of (\ref{Weyl-small-z}) corresponds to the Cotton tensor $C^{ij}=\frac{1}{\sqrt{g}}\epsilon^{ikl}\nabla_k\left(R^j_{l}-\frac{1}{4}\delta^j_l R\right)$ linearized around the 3-d flat metric, i.e.~$g_{ij}(\vec{x})=\delta_{ij}+\alpha_{ij}(\vec{x})$. Indeed, it is well-known that this is the tensor that is obtained by varying the gravitational Chern-Simons action with respect to the metric. The corresponding operator acting on $\alpha_{ij}$ in (\ref{Weyl-small-z}) is precisely the parity-odd part of the 3-d graviton kinetic term in (\ref{Kspin2}). Therefore, the mixed boundary conditions that correspond to gauging the spin 2 symmetry while adding the gravitational Chern-Simons action may be stated as
\begin{equation}
\frac{1}{2}\epsilon_{ikl} C_{kl z j}+i b_2 \frac{N}{k_2} C_{z i z j}\Big{|}_{z=0}=0\,,
\label{mixed-spin-2}
\end{equation}
where $b_2$ is a normalization factor. Introducing the self-dual and anti-self dual parts of the Weyl tensor
\begin{equation}
C_{\mu\nu\rho\sigma}^{\pm}=C_{\mu\nu\rho\sigma}\pm\frac{1}{2}\epsilon_{\mu\nu}^{\ \ \kappa\lambda} C_{\kappa\lambda\rho\sigma} \,,
\end{equation}
these may be also written as
\begin{equation}
e^{i\gamma_2} C^{+}_{zizj}\Big{|}_{z=0}=e^{-i\gamma_2}C^{-}_{zizj}\Big{|}_{z=0}\,,\qquad \quad
e^{i\gamma_2} = \sqrt{\frac{k_2+i c_2 N}{k_2-i c_2 N}}\,.
\label{spin2-Cpm}
\end{equation}

We can proceed with the higher-spin cases in analogy with the above discussion. For a higher-spin gauge field of spin $s$, there is a natural generalization of the Weyl curvature tensor which is constructed by taking up to $s$ space-time derivatives on the symmetric rank $s$ tensor $h_{\mu_1\ldots \mu_s}$. It corresponds to a tensor $C_{\mu_1\nu_1\mu_2\nu_2\cdots \mu_s\nu_s}$ with the symmetries of the two row Young tableaux, each row having length $s$; this is the $[s,s]$ representation of $SO(4)$.  As in the lower spin examples, it can be split into its self-dual and anti self-dual parts corresponding, in the two-component spinor notations, to the totally symmetric multispinors $C_{\alpha_1\cdots \alpha_{2s}}$ and $C_{\dot\alpha_1\cdots \dot\alpha_{2s}}$. This corresponds to the fact that the $[s,s]$ representation of $SO(4)$ splits into the sum $({\bf 2s+1},{\bf 1}) \oplus ({\bf 1},{\bf 2s+1})$ of $SU(2)\times SU(2)$ representations. Such HS Weyl tensors appear naturally in Vasiliev's formulation of the higher-spin gauge theory. They are contained in the master 0-form $B(x|y^{\alpha},\bar y^{\dot \alpha},z^{\alpha},\bar z^{\dot \alpha})$  (here $y,\bar y,z,\bar z$ denote the auxiliary twistor variables) as the components of degree $(2s,0)$ and $(0,2s)$ in $(y,\bar y)$ and independent of $z, \bar z$. By analogy with (\ref{spin1-Fpm}) and (\ref{spin2-Cpm}), we can state the general mixed boundary conditions for HS gauge fields in terms of the HS Weyl tensors as
\begin{equation}
e^{i\gamma_s} C^{+}_{zi_izi_2\cdots zi_s}\Big{|}_{z=0}=e^{-i\gamma_s}C^{-}_{zi_izi_2\cdots zi_s}\Big{|}_{z=0}\,,\qquad \quad
e^{i\gamma_s} = \sqrt{\frac{k_s+i b_s N}{k_s-i b_s N}}\,.
\label{spins-Cpm}
\end{equation}
These boundary conditions are expected to correspond to turning on the spin $s$ Chern-Simons term in the induced conformal HS theory at the boundary. As above, $k_s$ denotes the Chern-Simons coupling constant, and $b_s$ is a normalization factor. It would be interesting to explicitly derive the bulk-to-boundary propagators and Green's functions which solve (\ref{spin-s-EOM}) with boundary conditions (\ref{spins-Cpm})\footnote{The propagators derived e.g. in \cite{Giombi:2009wh} satisfy the ordinary $\Delta_{+}$ boundary conditions, corresponding to $\gamma_s=0$ in (\ref{spins-Cpm}).} and verify that they reproduce the structure of the two-point function of the boundary HS gauge field with Chern-Simons terms. It is likely that in the fully non-linear theory, where one may need to gauge all the boundary HS symmetries at once, the Chern-Simons couplings $k_s$ should be all related, leaving only one independent coupling. In Vasiliev's theory, where the HS Weyl curvatures are all contained into a single master form, it seems natural to impose the conditions (\ref{spins-Cpm}) for all spins by working at the level of the master form instead of its single components.

To compute from the bulk the change in free energy $\delta F^{(s)}= F_{\rm gauged}^{(s)}-F_{\rm free}^{(s)}$, where the gauged system includes the CS term, one would have to calculate the one-loop determinants in AdS$_4$ with the mixed boundary conditions (\ref{spins-Cpm}). We leave this for future work. Note, however, that the discussion of trivial gauge symmetries in the previous section still applies to the mixed boundary conditions. Since the boundary HS symmetry is gauged, we still need to use $\delta_{-}$ boundary conditions for the ghosts, leading to the same counting of trivial gauge transformations. This implies, as before, that $\delta F = \frac{1}{2} n_{s-1}\log N+O(N^0)$ at large $N$ with $N/k_s$ fixed. From this point of view, the square root structure
discussed in Section \ref{CS-bdy} should be recovered by  computing the $O(N^0)$ terms coming from the one-loop determinants. Namely, we can write (\ref{deltaFs}) as (recall $C\propto N$ and $W \propto k_s$)
\begin{equation}
\begin{aligned}
&\delta F^{(s)} = \frac{1}{2} n_{s-1} \log \left( \sqrt{C^2+W^2}\right)\\
&=\frac{1}{2} n_{s-1} \log N + \frac{1}{2} n_{s-1} \log \left(\sqrt{1+\left(\frac{k_s}{c_s N}\right)^2}\right)+\ldots \,.
\end{aligned}
\end{equation}
The second term, which is $O(N^0)$, should come from the evaluation of the bulk one-loop determinants with mixed boundary conditions. Note that this is in principle consistent with the structure of (\ref{spins-Cpm}), which depend on the ratio $k_s/N$ and not on $N$ and $k_s$ separately.

\section{Comments on half-integer spins}
\label{half-int}

So far our discussion has been restricted to the case where $J_s$ is a bosonic single-trace operator of integer spin $s$.
Of course, it is also possible to consider cases where $J_s$ is a fermionic single-trace operator of half-odd-integer spin; the double-trace operator
is still bosonic and can be added to the action. The simplest case of $s=1/2$ in $d=3$ has already been studied in the literature \cite{Allais:2010qq,Klebanov:2011gs}.
In this section we briefly consider generalizations of this result to higher half-integer spin.
As we have seen, the dual AdS$_{4}$ calculations tend to be simpler than the field theory calculations on $S^3$.  In this section we list some results obtained in the bulk, leaving comparisons with the explicit field theory calculations for future work.

Following~\cite{Camporesi:1992tm}
we see that in the half-integer spin case the spectral function is modified to
\es{spectral3Ferm}{
\mu(\lambda) = {\pi \lambda \over 16} \left[ \lambda^2 + \left( s + {1 \over 2} \right)^2 \right] \coth \pi \lambda \,.
}
With the operator $J_s$ a real fermion, the change $\delta F_\Delta^{(s)}$ acquires an additional minus sign compared to~\eqref{changeFPhi} because of the closed fermion loop.  We then find that for half-integer spin
\es{GenFormulaFerm}{
\delta F^{(s)}_\Delta = {(2 \, s + 1) \pi \over 6}  \int_{3/2}^\Delta \big( x - {3 \over 2} \big) (x + s - 1)(x - s - 2) \tan (\pi x) \,,
}
so that for arbitrary integer or half-integer spin we have the general formula
\es{superGenFormula}{
\delta F^{(s)}_\Delta = {(2 \, s + 1) \pi \over 6} (-1)^{2s} \int_{3/2}^\Delta \big( x - {3 \over 2} \big) (x + s - 1)(x - s - 2) \cot \big(\pi (x+s)\big) \,.
}
Note that for spin $1/2$ this agrees with the result in~\cite{Allais:2010qq,Klebanov:2011gs}.

We note that for $\Delta= s+1-\epsilon$ we find a logarithmic divergence of the form
\es{logdiv}{
\delta F^{(s)} = -{s(4s^2-1)\over 6} \log \epsilon \,.
}
This again suggests that for $\epsilon=0$, $\delta F^{(s)} = \frac 1 2 n^{d=3}_{s-1}  \log N$, where for $d=3$
\es{fermdegen}{
n^{d=3}_{s-1}= {s (4s^2-1)\over 3}= {(2s+1)!\over 3! (2s-2)!} \,.
}
This formula is the restriction to $d=3$ of (\ref{dimrep}).\footnote{For $d>3$ the formula (\ref{dimrep}) does not apply to half-integer $s$ because that formula was calculated with
$m_3=\ldots=0$, which does not make sense for spinors. It is plausible that we should instead consider the representations $m_1=m_2=s-1$
and $m_3=\ldots=1/2$. For example, for $d=5$ the dimension of the representation with $m_1=m_2=s-1$ and $m_3=1/2$ is
${(2s+3) (2s+2) (2s+1) (s-\frac 1 2)(s+\frac 1 2) (s+\frac 3 2)  (s+\frac 5 2) \over 3 \times 5!}$.
It would be interesting to check by a direct calculation that this gives the correct number of fermionic Killing tensors.}
As we have discussed, the logarithmic divergence in $\delta F^{(s)}$ appears for $s \geq 1$ and is associated with gauge transformations that act trivially on the spin $s$ gauge field.
For example, for $s=3/2$ such gauge transformations are simply the 4 Killing spinors in AdS$_4$.
More generally, for half-integer $s$, the Killing tensors transform in the $m_1=m_2=s-1$ spinor representation of $SO(4,1)$.
The counting of degeneracies of such representations is particularly simple because they are symmetric tensors of rank $2s-2$ with spinor indices.
Indeed, the formula (\ref{fermdegen}) is simply the number of such tensors where each index takes 4 values. We note that this applies to integer $s$ as well. Note also that (\ref{fermdegen}) precisely vanishes at $s=0$ and $s=1/2$, which correspond to the only cases in which we do not have gauge symmetries.

\section{Calculation of Weyl anomalies in even $d$}
\label{Weylsec}

In this section we discuss
an interesting application of alternate boundary conditions in AdS$_{d+1}$: we will show that they provide an efficient
method for finding the Weyl anomaly coefficients of conformal higher-spin field theories
in even dimensions $d$. In the $d=4$ case such theories were introduced in \cite{Fradkin:1985am}; an interacting conformal higher-spin theory including each spin
once was proposed in \cite{Segal:2002gd}.

For all $d$ the alternate boundary conditions in AdS$_{d+1}$ correspond to a theory where the dynamics of the spin $s$ gauge field is ``induced''
by its coupling to the conserved current $J_{\mu_1 \mu_2 \dots \mu_s}$. However, some properties of the theory depend significantly on whether $d$ is even or odd.
In odd $d$ the induced conformally invariant action is necessarily non-local as, for example, in 3-dimensional QED\@.
In even $d$ we instead find a local conformally invariant term multiplied by $\log (q^2/\Lambda^2)$.
Well-known examples of this in $d=4$ include $F_{\mu\nu}F^{\mu\nu}$ for $s=1$ and the Weyl tensor squared, $C_{\mu\nu\kappa\sigma}
C^{\mu\nu\kappa\sigma}$, for $s=2$. Their appearance is due to the structure of 2-point functions; for example,
\es{twopoint}{
\langle J_\mu (q) J_\nu (-q) \rangle\sim (q_\mu q_\nu - \delta_{\mu\nu} q^2) \log (q^2/\Lambda^2)\,.
}
The logarithmic term is due to the fact that in QED$_4$,
the quantum effects of the charged fields lead to a logarithmic flow of
the charge. Far in the IR the dynamics reduces to that of the free Maxwell field decoupled from the charged field. This is a conformal
field theory, and we will show how considering a massless gauge field in AdS$_5$ with alternate boundary conditions
gives the familiar anomaly coefficient $a_1=31/45$.\footnote{We recall that $a$ conventionally denotes the coefficient of the Euler density term in the Weyl anomaly. By our methods we do not have access to the $c$ coefficient, which is the one associated with the square of the Weyl tensor.} Similarly, for $s=2$ we will obtain $a_2=87/5$ in agreement
with the direct calculation~\cite{Fradkin:1983tg, Fradkin:1985am} in the conformal Weyl-squared gravity.\footnote{
The relation to the notation for anomaly coefficients used in \cite{Fradkin:1983tg} is $a=2\beta_2- 4 \beta_1$; see also \cite{Fradkin:1985am}.}

First, let us calculate the change in the Weyl anomaly coefficient produced by the double-trace flows with operators
$J_{\mu_1 \mu_2 \dots \mu_s} J^{\mu_1 \mu_2 \dots \mu_s}$, where
$J_{\mu_1 \mu_2 \dots \mu_s}$ is a spin $s$ single-trace operator of dimension $\Delta$, extending the earlier work of
\cite{Gubser:2002zh,Gubser:2002vv,Diaz:2007an}.
When $d$ is even, the $\log{R}$ term in the free energy on $S^d$ is identified with the anomaly $a$-coefficient.
Using \eqref{zeta1GEN} we then find
\es{finalEven}{
\delta a^{(s)}_\Delta =-&{2 g(s) \over \pi \, d!} \int_{\frac{d}{2}}^{\Delta}dx\left( x - {d \over 2} \right) (x + s -1 ) (x -s -d +1)
\Gamma(x -1) \Gamma(d-1 - x) \sin(\pi x) \,.
}
For $s=0$ this expression agrees with the results in~\cite{Gubser:2002zh,Gubser:2002vv,Diaz:2007an}. With $s=0, \Delta= \frac{d}{2}+1$ this formula agrees with the coefficient of the logarithmic divergence in the $S^d$ free energy for a conformally coupled scalar field \cite{Casini:2010kt}. For instance, $\delta a^{(0)}=-\frac{1}{3},\frac{1}{90},-\frac{1}{756},\frac{23}{113400}$ in $d=2,4,6,8$ respectively. This is because in this case the Hubbard-Stratonovich field has the dimension of a free conformal scalar.

An interesting special case is $d = 4$.  Integrating~\eqref{finalEven} over $\Delta$ we obtain the change in the $a$-anomaly coefficient:
\es{deltaA}{
\delta a^{(s)}= a^{(s)}_\text{UV} - a^{(s)}_\text{IR} =  {(s+1)^2 \over 180} (\Delta - 2)^3 \big[ 5 (1 + s)^2 - 3 (\Delta -2)^2 \big] \,,
}
where $a$ is normalized such that $a = 1/ 90$ for a real conformal scalar field. 

The higher-spin conformal gauge theories are obtained by taking $\Delta = 2 + s$ with $s \geq 1$, but in this case we must be careful to also include the contribution of the spin $s-1$ ghosts with alternate boundary conditions.  Since the ghost determinant appears in the numerator of~\eqref{one-loop-gauge}, the contribution of the ghosts to the anomaly $a$-coefficient of the induced theory may be computed from~\eqref{deltaA} with $\Delta = 3 + s$ (recall that for the spin $s-1$ ghosts we have $\Delta_{\pm}=\delta_{\pm}+s-1$, where $\delta_{\pm}$ is given in (\ref{ghost-bc})).  More explicitly, defining $a_s= a_s^\text{gauged} - a_s^\text{ungauged}$ so that $a_s$ is the anomaly $a$-coefficient for the conformal spin $s$
field, we have
\es{gaugedungauged}{
a_s = a_s^\text{phys} - a_{s-1}^\text{ghost}\ ,
}
with $a_s^\text{phys}$ the contribution from the physical modes and $ a_{s-1}^\text{ghost}$ that from the ghosts. We find
\es{aPG}{
a_s^\text{phys} &= {s^3 \over 180}  (1 + s)^2 \big[ 5+ 2\, s \,(5 + s) \big] \,, \\
a_{s-1}^\text{ghost} &= - {s^2 \over 180} (1+s)^3 \big[ 3 + 2\, s\, ( 3 - s) \big] \,,
}
which leads to the result quoted in~\eqref{higheranom}. Using this result, we can calculate the Weyl anomaly of the 4-d conformal gauge theory including
the fields of each positive integer spin once. One way to try constructing such an induced gauge theory is to start with $N$ conformal
charged scalars or fermions in $d=4$ and gauge all the currents with $s\geq 1$. Using \eqref{higheranom} and the zeta-function regularization, we find that the sum of all
Weyl anomaly coefficients
\es{allspins}{
\sum_{s=1}^\infty a_s= \frac {1} {90} \big[ 10 \zeta(-3)+ 21 \zeta(-5) \big]= 0
\ ,
}
where we have used the fact that $\zeta(-2n)=0$ for $n>1$. Thus, the theory with such a field content has no $a$-type Weyl anomaly. This provides
partial evidence for
the consistency of such a conformal higher-spin theory, but the $c$ anomaly coefficient remains to be determined.

Since the $a$-type Weyl anomaly cancels in the conformal higher-spin theory, the leading term in the $S^4$ free energy of the induced theory is the $\log N$ type term that comes from~\eqref{deltaF-logN}.  When the $a$-type anomaly does not cancel in an even dimensional induced gauge theory, this term is subdominant compared to the $\log R$ term.  The sum over all of the $\log N$ contributions in zeta-function regularization gives
\es{sumF}{
F = {1 \over 2} \sum_{s = 1}^\infty n_{s-1} \log N = {\log N \over 24} \big( \zeta(-2) + 4 \zeta(-3) + 5 \zeta(-4) + 2 \zeta(-5) \big) = {\log N \over 945} \,,
}
where we have used~\eqref{dimrep} to calculate $n_{s-1}$ in $d=4$.

A similar calculation may be carried out in other even dimensions; for example, in $d = 2$ we find that for generic $\Delta$ the change in central charge is given by
\es{deltaC}{
c_\text{UV} - c_\text{IR} = g(s) (\Delta - 1)\big[ (\Delta - 1)^2 - 3 s^2 \big]
}
in units where $c = 1$ for a real scalar field.  When the dimension $\Delta$ equals the spin so that we are dealing with a spin $s$ gauge theory,
we may include the contribution of the ghosts to calculate
$c_s= c_s^\text{gauged} - c_s^\text{ungauged}$. We find that
\es{Cs}{
c_1 = - 1 \,, \qquad c_s = - 2 \big[ 1 + 6 \, s\, (s - 1) \big]  \quad (s \geq 2) \,.
}
The central charges $c_s$ with $s \geq 2$ agree with those in the $W$-gravity theories \cite{Pope:1991uz}; they are the central charges of the higher-spin $bc$
 ghost system with weights $(s,1-s)$.
 In particular, for
 $s = 2$ we find the well-known result $c_2 = -26$ for the central charge of the ghost system in the 2-d gravity \cite{Polyakov:1981rd}.
 Thus, we have found a dual AdS$_3$ approach to the critical dimension of the bosonic string. We note that the result for $s=2$
 does not include the contribution of the conformal factor, the Liouville mode. This mode is frozen because in the dual AdS$_3$ calculation the trace of the graviton at the boundary is kept fixed to zero. Similarly, in the calculation of the Weyl anomaly for 4-d conformal gravity the conformal factor is frozen. The
 result $a_2=87/5$ of~\cite{Fradkin:1983tg, Fradkin:1985am} is obtained in a ``quantum Weyl gauge," where the trace of the graviton is set to zero off-shell, and so $a_2$ receives contributions only from the traceless gravitons and ghosts.\footnote{Of course, in the presence of a net non-zero anomaly, the conformal factor does not really decouple and becomes dynamical, as in the quantum Liouville theory \cite{Polyakov:1981rd}. But the result $a_2=87/5$ does not include the contribution of this trace mode.}

As noted in~\cite{Pope:1991uz}, in zeta function regularization
\es{sumcs}{
\sum_{s= 2}^\infty c_s =  2 \big[ 1 +6 \zeta(-1)  - \zeta(0) \big] = 2 \,.
}
Thus, a conformal 2-d theory with $s\geq 2$ fields does not have a vanishing Weyl anomaly. However, as observed in~\cite{Pope:1991uz}, it is possible to cancel the total anomaly by adding a suitable matter sector with $c_{\rm mat}=-2$. A well-known example is the ``topological'' $\eta \xi$ theory with weights $(1,0)$;
it is the $s=1$ case of the $bc$ ghost systems with weights $(s,1-s)$.

\section*{Acknowledgments}
We thank J. Maldacena, B. Nilsson, A. Tseytlin, M. Vasiliev, E. Witten and X. Yin for helpful comments and discussions. The work of IRK, BRS, and GT was supported in part by the US NSF under Grants No.~PHY-0756966 and PHY-1314198.  SSP was supported in part by a Pappalardo Fellowship in Physics at MIT and in part by the U.S. Department of Energy under cooperative research agreement Contract Number DE-FG02-05ER41360\@.   IRK thanks the Galileo Galilei Institute for Theoretical Physics for the hospitality and the INFN for partial support during his work on this paper. IRK also thanks the organizers of Strings 2013 conference for the hospitality and an opportunity to present this work.

\appendix

\section{Symmetric traceless tensor harmonics on $S^3$}
\label{HARMONICSAPP}

In this Appendix we collect a few useful results on $S^3$ tensor spherical harmonics.  Most of these results can also be found elsewhere in the literature---see, for instance, \cite{Tomita:1982ew, Rubin:1983be, Higuchi:1986wu, Jantzen}.   For presenting explicit formulas for the tensor harmonics, it is convenient to use the standard coordinates $(\chi, \theta, \phi)$ on the three-sphere, for which the line element takes the form
 \es{LineElement}{
  ds^2 = d\chi^2 + \sin^2 \chi \left(d \theta^2 + \sin^2 \theta d\phi^2 \right) \,.
 }
The angles $(\theta, \phi)$ are the standard coordinates on an equatorial $S^2$, whose $SO(3)$ isometry group embeds diagonally into the isometry group $SO(4) \cong SU(2)_L \times SU(2)_R$ of the three-sphere.

As described in Section~\ref{HARMONICS}, the Hilbert space of normalizable traceless symmetric tensors of rank-$s$ decomposes under $SU(2)_L \times SU(2)_R$ as
  \es{IrredAgain}{
  \bigoplus_{n=s+1}^\infty \bigoplus_{s' = -s}^s ({\bf n+ s'}, {\bf n - s'} ) \,.
 }
We denoted the basis of tensors in the $({\bf n+ s'}, {\bf n - s'} )$ by $\HH^{s', n\ell m}_{\mu_1 \ldots \mu_s}(x)$, where $\abs{s'} \leq  \ell < n$ and $m = -\ell, -\ell+1, \ldots, \ell$.  Group theory implies that \cite{Higuchi:1986wu, Jantzen}
 \es{LapTensor}{
  \nabla^\nu \nabla_\nu \HH^{s', n\ell m}_{\mu_1 \ldots \mu_s} &=
    -\left(n^2 + s'^2 - 1 - s(s+1) \right) \HH^{s', n\ell m}_{\mu_1 \ldots \mu_s} \,, \\
  \nabla^\nu \HH^{s', n\ell m}_{\nu \mu_1 \ldots \mu_{s-1}} &=  -\sqrt{\frac{(n^2 - s^2)(s^2 - s'^2)}{s (2s-1)}} \, \HH^{s', n\ell m}_{\mu_1 \ldots \mu_{s-1}} \,.
 }
When $s'<s$, one can construct the tensor harmonics $\HH^{s', n\ell m}_{\mu_1 \ldots \mu_s}(x)$ recursively from harmonics of lower rank:
\es{Recursion}{
   \HH^{s', n\ell m}_{\mu_1 \ldots \mu_s} =\sqrt{\frac{s (2s-1)}{(n^2 - s^2)(s^2 - s'^2)}} \left[ \nabla_{(\mu_1} \HH^{s', n\ell m}_{\mu_2 \mu_3 \dots \mu_s)} - \frac{s-1}{2s-1} g_{(\mu_1 \mu_2 }\nabla^\nu \HH^{s', n\ell m}_{\mu_3 \mu_4 \dots \mu_{s}) \nu} \right]\,,
}
where the overall normalization is fixed by requiring $\HH^{s', n\ell m}_{\mu_1 \ldots \mu_s}$ to have unit norm, namely
 \es{UnitNorm}{
  \int \sin^2 \chi \sin \theta\, d\chi\, d\theta\, d\phi\, \HH^{s', n\ell m}_{\mu_1 \ldots \mu_s}(\chi, \theta, \phi)^*\, \HH_{s', n\ell m}^{\mu_1 \ldots \mu_s}(\chi, \theta, \phi)  = 1 \,.
 }
In \eqref{Recursion}, one recognizes the operator ${\cal O}_g$ defined in \eqref{Og} acting on a rank-$(s-1)$ tensor.  All the tensors $\HH^{s', n\ell m}_{\mu_1 \ldots \mu_s}$ can therefore be straightforwardly constructed from knowing those with $s' = s$ for all $s$.  These latter tensors are covariantly conserved, as \eqref{LapTensor} reduces in this case to $\nabla^\nu \HH^{s, n\ell m}_{\nu \mu_1 \ldots \mu_{s-1}} = 0$.

The formulas we are about to present simplify if we also make use of the $\Z_2$ parity symmetry, which acts by interchanging $SU(2)_L$ with $SU(2)_R$, so it sends the $({\bf n + s'}, {\bf n - s'})$ representation to $({\bf n - s'}, {\bf n + s'})$.  For $s'>0$, it is convenient to define the odd and even combinations
 \es{HOddEven}{
  \EE^{s', n \ell m}_{\mu_1 \ldots \mu_s}(x) &= \frac{1}{\sqrt{2}} \left( \HH^{s', n\ell m}_{\mu_1 \ldots \mu_s} (x)+ \HH^{-s', n\ell m}_{\mu_1 \ldots \mu_s} (x)\right) \,, \\
  \OO^{s', n \ell m}_{\mu_1 \ldots \mu_s}(x) &= \frac{1}{\sqrt{2}} \left( \HH^{s', n\ell m}_{\mu_1 \ldots \mu_s} (x) - \HH^{-s', n\ell m}_{\mu_1 \ldots \mu_s} (x) \right)  \,,
 }
where the normalization is such that if $\HH^{s', n\ell m}_{\mu_1 \ldots \mu_s}$ has unit norm, then so do $\EE^{s', n\ell m}_{\mu_1 \ldots \mu_s}$ and $\OO^{s', n\ell m}_{\mu_1 \ldots \mu_s}$.  For $s'=0$ we can take  $\EE^{0, n \ell m}_{\mu_1 \ldots \mu_s}(x) = \HH^{0, n \ell m}_{\mu_1 \ldots \mu_s}(x)$.

Since the kernels in Section~\ref{HARMONICS} are parity-even, the only harmonics that will be relevant are the even ones.  Indeed, one can define
 \es{ZE}{
  \Z^{(E) s', n}_{\mu_1 \ldots \mu_s ; \nu_1 \ldots \nu_s}(x)
   = \sum_{\ell, m} \EE^{s', n \ell m}_{\mu_1 \ldots \mu_s}(x)^* \EE^{s', n \ell m}_{\nu_1 \ldots \nu_s} (0) \,,\\
  \Z^{(O) s', n}_{\mu_1 \ldots \mu_s ; \nu_1 \ldots \nu_s}(x)
   = \sum_{\ell, m} \OO^{s', n \ell m}_{\mu_1 \ldots \mu_s}(x)^* \OO^{s', n \ell m}_{\nu_1 \ldots \nu_s} (0) \,,
 }
and write
 \es{ZRewrite}{
  \Z^{s', n}_{\mu_1 \ldots \mu_s ; \nu_1 \ldots \nu_s}(x)
   + \Z^{-s', n}_{\mu_1 \ldots \mu_s ; \nu_1 \ldots \nu_s}(x)
   = \Z^{(E) s', n}_{\mu_1 \ldots \mu_s ; \nu_1 \ldots \nu_s}(x) + \Z^{(O) s', n}_{\mu_1 \ldots \mu_s ; \nu_1 \ldots \nu_s}(x) \,.
 }
It can be checked that $\OO^{s', n\ell m}_{\mu_1 \ldots \mu_s} (0) = 0$ and hence $\Z^{(O) s', n}_{\mu_1 \ldots \mu_s ; \nu_1 \ldots \nu_s}(x) = 0$.  We then have
 \es{ZspZE}{
  \Z^{(E) s', n}_{\mu_1 \ldots \mu_s ; \nu_1 \ldots \nu_s}(x)
   = \begin{cases}
    \Z^{s', n}_{\mu_1 \ldots \mu_s ; \nu_1 \ldots \nu_s}(x)
   + \Z^{-s', n}_{\mu_1 \ldots \mu_s ; \nu_1 \ldots \nu_s}(x)\,, & \text{if $s' >0$} \,, \\
   \Z^{0, n}_{\mu_1 \ldots \mu_s ; \nu_1 \ldots \nu_s}(x)\,, & \text{if $s'=0$} \,.
   \end{cases}
 }
Using \eqref{knSimpAgain} as well as the fact that for a parity-invariant theory $k_{n, s'} = k_{n, -s'}$, we can then use
 \es{knEven}{
  k_{n, s'} = k_{n, -s'} = \frac{32 \pi^3}{n^2 - s'^2} \int dr \, \frac{r^2}{\left(1 + r^2 \right)^3}\, \Z^{(E) s', n}_{\mu_1\ldots \mu_s; \nu_1 \ldots \nu_s}(r \hat v) \,
    K^{\mu_1\ldots \mu_s; \nu_1 \ldots \nu_s}(r \hat v, 0)
 }
instead of \eqref{knSimpAgain} whenever $s' > 0$.  When $s'=0$ we can still use \eqref{knSimpAgain}.

\subsection{Even harmonics}

\subsubsection{Spin $0$} \label{spin0}

For $s = 0$, we have $\EE^{0, n \ell m}(\chi, \theta, \phi) = \HH^{0, n \ell m}(\chi, \theta, \phi) = Y_{n \ell m}(\chi, \theta, \phi)$.  An explicit expression can be found by writing
 \es{Ynlm}{
  Y_{n \ell m}(\chi, \theta, \phi) = \sin^\ell \chi\, \Phi_{n \ell}(\chi) Y_{\ell m}(\theta, \phi) \,,
 }
where $Y_{\ell m}(\theta, \phi)$ are the $S^2$ spherical harmonics and then using \eqref{LapTensor} to find a second order differential equation satisfied by $\Phi_{n \ell}(\chi)$.  The regular solutions of this equations are
 \es{PhiReg}{
  \Phi_{n \ell}(\chi) =\frac{1}{\sqrt{a_{n \ell}}} \frac{d^{\ell+1} \cos (n \chi)}{d(\cos \chi)^{\ell+1}} \,, \qquad a_{n \ell} =  \frac{n \pi (\ell + n)!}{2 (n - \ell - 1)!} \,.
 }
The normalization factor in \eqref{PhiReg} was chosen so that $Y_{n \ell m}(\chi, \theta, \phi)$ has unit norm on $S^3$.

\subsubsection{Spin $1$} \label{spin1}

The spin-$1$ even harmonics can be expanded as
 \es{Spin1Expansion}{
  \EE^{s', n \ell m}_\chi (\chi, \theta, \phi) &= V_1^{n \ell}(\chi) Y_{\ell m}(\theta, \phi) \,, \\
  \EE^{s', n \ell m}_\alpha(\chi, \theta, \phi) &=  V_2^{n \ell}(\chi) \sin \chi\, \hat \nabla_\alpha Y_{\ell m}(\theta, \phi) \,,
 }
where $\alpha = \theta, \phi$ and $\hat \nabla_\alpha$ is the covariant derivative on $S^2$.  Eqs.~\eqref{LapTensor} uniquely determine $V_1^{n \ell}(\chi)$ and $V_2^{n \ell}(\chi)$ up to an overall normalization.

For $s' = 0$ we have
 \es{Vsp0}{
  V_1^{n \ell} (\chi) &= \frac{1}{\sqrt{a_{n \ell} (n^2 - 1)}} \frac{d}{d \chi} \left( \sin^\ell \chi  \frac{d^{\ell+1} \cos (n \chi)}{d(\cos \chi)^{\ell+1}} \right)  \,, \\
  V_2^{n \ell} (\chi) &= \frac{1}{\sqrt{a_{n \ell} (n^2 - 1)}} \sin^{\ell -1 } \chi \frac{d^{\ell+1} \cos (n \chi)}{d(\cos \chi)^{\ell+1}} \,,
 }
which follows from either solving \eqref{LapTensor} or from combining the recursion relation \eqref{Recursion} with \eqref{Ynlm}.  For $s'=1$, solving \eqref{LapTensor} yields
 \es{Vsp1}{
  V_1^{n \ell} (\chi) &= \frac{\sqrt{\ell(\ell+1)}}{n \sqrt{a_{n \ell} }} \sin^{\ell-1} \chi \frac{d^{\ell+1} \cos (n \chi)}{d(\cos \chi)^{\ell+1}} \,, \\
  V_2^{n \ell} (\chi) &= \frac{1}{n \sqrt{a_{n \ell} \ell(\ell+1) }} \frac{1}{\sin \chi} \frac{d}{d\chi} \left(\sin^{\ell+1} \chi \frac{d^{\ell+1} \cos (n \chi)}{d(\cos \chi)^{\ell+1}} \right)  \,.
 }

\subsubsection{Spin $2$}

The spin-$2$ even harmonics can be expanded as
 \es{Spin2Expansion}{
  \EE^{s', n \ell m}_{\chi \chi} (\chi, \theta, \phi) &= T_1^{n \ell}(\chi) Y_{\ell m}(\theta, \phi) \,, \\
  \EE^{s', n \ell m}_{\chi \alpha} (\chi, \theta, \phi) &=  T_2^{n \ell}(\chi) \sin \chi\, \hat \nabla_\alpha Y_{\ell m}(\theta, \phi) \,, \\
  \EE^{s', n \ell m}_{\alpha \beta} (\chi, \theta, \phi) &=  \sin^2 \chi\,\left[ T_3^{n \ell}(\chi)
    \hat \nabla_\alpha \hat \nabla_\beta Y_{\ell m}(\theta, \phi)
   + T_4^{n \ell}(\chi)  \hat g_{\alpha \beta} Y_{\ell m}(\theta, \phi) \right] \,.
 }
For $s'<2$ one can use the recursion relation \eqref{Recursion} to find explicit expressions for $T_i^{n \ell}$, which we will not reproduce here.  For $s' = 2$, one can solve the equations \eqref{LapTensor}, whose normalized solutions are \cite{Tomita:1982ew}
 \es{T2}{
  T_1^{n\ell} &= \sqrt{ \frac{(\ell+2)! (n-2)!}{2n a_{n \ell} (\ell-2)! (n+1)! } } \sin^{\ell-2} \chi \frac{d^{\ell+1} \cos (n \chi)}{d(\cos \chi)^{\ell+1}} \,, \\
  T_2^{n \ell} &= \frac{1}{\ell(\ell+1)} \left[ \sin \chi (T_1^{n \ell})' + 3 \cos \chi T_1^{n \ell} \right]  \,, \\
  T_3^{n \ell} &= \frac{1}{(\ell-1)(\ell+2)} \left[2 \sin \chi (T_2^{n \ell})' + 6 \cos \chi T_2^{n \ell} - T_1^{n \ell} \right] \,, \\
  T_4^{n \ell} &= \frac 12 \left[\ell(\ell+1) T_3^{n\ell} - T_1^{n \ell} \right] \,.
 }

\subsubsection{Spin $3$}

The spin-$3$ even harmonics can be expanded as
 \es{Spin3Expansion}{
  \EE^{s', n \ell m}_{\chi \chi \chi} (\chi, \theta, \phi) &= U_1^{n \ell}(\chi) Y_{\ell m}(\theta, \phi) \,, \\
  \EE^{s', n \ell m}_{\chi \chi \alpha} (\chi, \theta, \phi) &=  U_2^{n \ell}(\chi) \sin \chi\, \hat \nabla_\alpha Y_{\ell m}(\theta, \phi) \,, \\
  \EE^{s', n \ell m}_{\chi \alpha \beta} (\chi, \theta, \phi) &=  \sin^2 \chi\,\left[ U_3^{n \ell}(\chi)
    \hat \nabla_\alpha \hat \nabla_\beta Y_{\ell m}(\theta, \phi)
   + U_4^{n \ell}(\chi)  \hat g_{\alpha \beta} Y_{\ell m}(\theta, \phi) \right] \,, \\
  \EE^{s', n \ell m}_{\alpha \beta \gamma} (\chi, \theta, \phi) &=  \sin^3 \chi\,\left[ U_5^{n \ell}(\chi)
    \hat \nabla_{(\alpha} \hat \nabla_\beta \hat \nabla_{\gamma)} Y_{\ell m}(\theta, \phi)
   + U_6^{n \ell}(\chi)  \hat g_{(\alpha \beta} \hat \nabla_{\gamma)} Y_{\ell m}(\theta, \phi) \right] \,.
 }
For $s'<3$ one can find the $U_i^{n \ell}$ by using the recursion formula \eqref{Recursion}.  For $s' = 3$, the solution of \eqref{LapTensor} is
 \es{URules3}{
  U^{n \ell}_1 &= \sqrt{\frac{(\ell+3)! (n-3)! }{4n a_{n \ell}(\ell -3)! (n+2)! }} \sin^{\ell - 3} \frac{d^{\ell+1} \cos (n \chi)}{d(\cos \chi)^{\ell+1}} \,, \\
  U^{n \ell}_2 &= \frac{1}{\ell (\ell+1)} \left[4 \cos \chi U^{n \ell}_1 + \sin \chi (U_1^{n \ell})' \right] \,, \\
  U^{n\ell}_3 &= \frac{1}{(\ell-1)(\ell+2)} \left[2 \sin \chi (U_2^{n \ell})' + 8 \cos \chi U_2^{n \ell} - U_1^{n \ell} \right] \,, \\
  U^{n \ell}_4 &= \frac 12 \left[\ell(\ell+1) U^{n \ell}_3 - U^{n \ell}_1 \right] \,, \\
  U_5^{n \ell} &= \frac{1}{(\ell-2)(\ell+3)} \left[2 \sin \chi (U_3^{n \ell})' + 8 \cos \chi U_3^{n \ell} - U_2^{n \ell} \right] \,, \\
  U^{n \ell}_6 &= \frac{1}{4} \left[ (3 \ell^2 + 3 \ell -2) U^{n \ell}_5 - 3 U^{n \ell}_2 \right] \,.
 }

\section{Eigenvalues of the integration kernel}
\label{EVALUESAPP}

\subsection{Spin $1$} \label{spin1Eigen}

When using \eqref{knSimpAgain} and \eqref{knEven} to compute the eigenvalues $k_{n, s'}$ of the kernel $K$, it is more convenient to use the frame
 \es{FrameS3}{
  \hat e^1 &= d \chi \,, \\
  \hat e^2 &= \sin \chi d\theta \,, \\
  \hat e^3 &= \sin \chi \sin \theta d\phi \,,
 }
which is different from the frame \eqref{S3Frame} introduced earlier.  In the frame \eqref{FrameS3}, the kernel $K$ takes the form
 \es{KFrame}{
  K_{\hat i \hat j}(r \hat v, 0) = \frac{1}{4^\Delta \sin (\chi /2)^{2 \Delta}} \diag \{-1, 1, 1\} \,,
 }
where we wrote $r = \tan (\chi/2)$ as in \eqref{rDef}.  In the same frame, using the results of section~\ref{spin1} we can write
 \es{ZonalV}{
  \Z^{(E) s', n}_{\hat i \hat j}(\chi, 0, 0) = \frac{3}{4 \pi} \diag\{V_1^{n1}(\chi) V_1^{n1}(0) , V_2^{n1} (\chi) V_2^{n1}(0),  V_2^{n1}(\chi) V_2^{n1}(0)\} \,.
 }
This expression holds for both $s' = 0$ and $s' = 1$ with the functions $V_i^{n \ell}$ defined in \eqref{Vsp0} and \eqref{Vsp1}, respectively.

Combining \eqref{KFrame} and \eqref{ZonalV} with \eqref{knSimpAgain} (for $s'=0$) and \eqref{knEven} (for $s'=\pm 1$), we obtain
 \es{ks1}{
  k_{n, s'} = \frac{2 \pi N C}{4^{\Delta} n (n^{2}-1)} \int_{0}^{\pi} d\chi \frac{f_{n, s'}(\chi) \sin \chi }{\sin (\chi/2)^{2+2\Delta}} \,,
 }
where
 \es{Gotf1}{
   f_{n, 0}(\chi) &=
   (1-n^{2}+\cos \chi (n^{2}+1))\sin n\chi -2n \sin \chi  \cos n\chi \,, \\
   f_{n, \pm 1}(\chi) &=
    (n^{2}-n^{2}\cos \chi -1)\sin n\chi + n \sin \chi \cos n\chi\,.
 }
With the help of the integrals
 \es{int}{
  I_{n}^{\Delta} = \frac{1}{n}\int_{0}^{\pi}d\chi \frac{ \sin \chi \sin n\chi}{\sin (\chi/2)^{2+2\Delta}} &= -\frac{4^{\Delta + 1} \sin (\pi \Delta) \Gamma(-2\Delta)\Gamma(n+\Delta)}{\Gamma(1+n-\Delta)} \,, \\
  \int_{0}^{\pi}d\chi  \frac{ \sin^{2} \chi \cos n\chi}{\sin (\chi/2)^{2+2\Delta}}  &= \frac{1}{2}\left[ (n+1)I_{n+1}^{\Delta}-(n-1)I_{n-1}^{\Delta} \right] \,, \\
   \int_{0}^{\pi}d\chi  \frac{ \sin \chi \cos \chi \sin n \chi}{\sin(\chi/2)^{2+2\Delta}} &= \frac{1}{2} \left[ (n+1)I_{n+1}^{\Delta}+(n-1)I_{n-1}^{\Delta} \right]  \,,
 }
one immediately finds the result quoted in \eqref{results1}.

\subsection{Spin $2$}

In this case we find
 \es{kn2}{
  k_{n, s'} =  \frac{2 \pi N C}{4^{\Delta+1} n (n^{2}-1) (n^2 - 4)} \int_{0}^{\pi} d\chi \frac{f_{n, s'}(\chi) \sin \chi }{\sin (\chi/2)^{4+2\Delta}} \,,
 }
where
 \es{Gotf2}{
  f_{n, 0}(\chi) &= -12 n \left [4 - n^2 + (2 + n^2) \cos \chi \right] \sin \chi \cos n \chi   \\
     &+ \left[3 (12 - 7n^2 + n^4) + 4 (8 +2n^2 - n^4) \cos \chi + (4 + 13n^2 + n^4) \cos 2 \chi \right] \sin n \chi \,, \\
  f_{n, \pm 1}(\chi) &= 2n \left[5(4 - n^2) + (4 + 5n^2) \cos \chi \right]   \sin \chi \cos n \chi \\
    &- \left[3 (8 - 6n^2 + n^4) + (24 + 10n^2 - 4n^4) \cos \chi + n^2 (8 + n^2) \cos 2 \chi \right] \sin n \chi \,, \\
  f_{n, \pm 2}(\chi) &= -4n \left[4 - n^2 + (-1 + n^2) \cos \chi  \right]  \sin \chi \cos n \chi \\
   &+ \left[4n^2(4 - n^2) \cos \chi + (1 - n^2) (12 - 3n^2 - n^2 \cos 2 \chi) \right] \sin n \chi \,.
 }
Performing the integrals in \eqref{kn2} explicitly, one then finds the expressions in \eqref{svtTens}.

\subsection{Spin $3$}
\label{SPIN3}

Lastly, for $s = 3$ we have
 \es{kn3}{
  k_{n, s'} =  \frac{2 \pi N C}{4^{\Delta+1} n (n^{2}-1) (n^2 - 4)(n^2 - 9)} \int_{0}^{\pi} d\chi\, \frac{f_{n, s'}(\chi) \sin \chi }{\sin (\chi/2)^{6+2\Delta}} \,,
 }
where
 \es{Gotf21}{
  f_{n, 0}(\chi) &= -12 n \biggl[  76 - 23 n^2 + n^4 + (63 + 11n^2 - 2 n^4) \cos \chi \\
   &+ (11 + 12 n^2 + n^4)\cos^2 \chi \biggr] \sin \chi \cos n \chi \\
   &+ \biggl[ 576 - 649 n^2 + 74 n^4 - n^6 + 3(288 - 43 n^2 - 30 n^4 + n^6) \cos \chi \\
   &+ (324 + 585 n^2 - 42 n^4 - 3n^6)\cos^2 \chi
    + (36 + 193 n^2 + 58 n^4 + n^6) \cos^3 \chi   \biggr] \sin n \chi \,, }
 \es{Gotf22}{
 f_{n, \pm 1}(\chi) &= n \biggl[756  - 233 n^2 + 11 n^4 + (558 + 136 n^2 - 22n^4) \cos \chi \\
   &+ (36 + 97n^2 + 11n^4) \cos^2 \chi \biggr]  \sin \chi \cos n \chi \\
   &+ \biggl[  -486 + 549 n^2 - 64 n^4 + n^6 + (-648 + 36n^2 + 81 n^4 - 3n^6) \cos \chi \\
   &+ 3(-72 -163n^2 + 10n^4 + n^6) \cos^2 \chi - n^2 (96 + 47n^2 + n^4) \cos^3 \chi \biggr]  \sin n \chi \,, }
 \es{Gotf23}{
 f_{n, \pm 2}(\chi) &= -2n \biggl[ 216 - 70 n^2 + 4n^4 + (63 + 65n^2 - 8n^4) \cos \chi \\
   &+ (-9 + 5n^2 + 4n^4) \cos^2 \chi \biggr]   \sin \chi \cos n \chi \\
   &+\biggl[ 270 - 309 n^2 + 40 n^4 - n^6 + 3 (90 + 29 n^2 - 20 n^4 + n^6) \cos \chi \\
   &+3n^2 (81 - n^4) \cos^2 \chi + n^2 (-21 + 20n^2 + n^4) \cos^3 \chi \biggr] \sin n \chi \,, }
 \es{Gotf24}{
 f_{n, \pm 3}(\chi) &= 3n \biggl[ 44 - 15 n^2 + n^4 - 2 (9 - 10n^2 + n^4) \cos \chi \\
   &+ (4 - 5n^2 + n^4) \cos^2 \chi \biggr]   \sin \chi \cos n \chi \\
   &+\biggl[-90 + 109 n^2 - 20 n^4 + n^6 - 3n^2 (44 - 15 n^2 + n^4) \cos \chi  \\
   &+3n^2 (9 - 10n^2 + n^4) \cos^2 \chi - n^2 (4 - 5n^2 + n^4) \cos^3 \chi \biggr] \sin n \chi \,.
 }
Performing these integrals, we obtain
 \es{k3Answer}{
  k_{n, 0} &= c(\Delta) \frac{\Gamma(n-1+\Delta)}{\Gamma(n+2-\Delta)} \,, \qquad
   k_{n, \pm 1} = \frac{1 - \Delta}{\Delta - 2} k_{n, 0}  \,, \\
  k_{n, \pm 2} &= \frac{\Delta(\Delta-1)}{(\Delta-2)(\Delta-3)} k_{n, 0} \,, \qquad
   k_{n, \pm 3} = -\frac{(\Delta+1)\Delta (\Delta-1)}{(\Delta-2)(\Delta-3)(\Delta-4)} k_{n, 0}  \,,
 }
where
 \es{Gotc3}{
  c(\Delta) = -\frac{4 NC (\Delta-2)(\Delta-3)(\Delta-4) \Gamma(2 - 2 \Delta) \sin(\pi \Delta)}{\Delta (\Delta + 1)(\Delta + 2)} \,.
 }
These expressions are consistent with the general conjecture \eqref{snvnArb}.

\bibliographystyle{ssg}
\bibliography{CGLP}

\end{document}